\documentclass[manuscript, nonacm]{acmart}

\AtBeginDocument{%
  }

\usepackage{tikz}
\usepackage{tcolorbox}
\usepackage{graphicx} 
\usepackage{subcaption}
\usepackage{soul}
\usepackage{tabularx}
\usepackage{array}
\usepackage{xltabular}
\usepackage[pdftex]{changebar}
\usepackage{cleveref}

\usepackage{wrapfig}

\usepackage{lipsum}
\usepackage{todonotes}
\usepackage{xcolor}
\usepackage{algorithm,algpseudocode}
\usepackage{amsmath}
\usepackage{bbm}
\usepackage{wrapfig}
\usepackage{booktabs,multirow,array}
\usepackage{colortbl}
\usepackage{makecell} 

\definecolor{myLocal}{HTML}{d95f02}
\definecolor{myGlobal}{HTML}{1b9e77}
\definecolor{myAxis}{HTML}{7570b3}

\definecolor{colorConfigurationLogic}{HTML}{f39882}
\definecolor{colorWorldModel}{HTML}{fddbaa}
\definecolor{colorObservation}{HTML}{d2e7ca}
\definecolor{colorCommunication}{HTML}{9ed1ea}
\definecolor{colorActions}{HTML}{e0c2dc}
\definecolor{colorInfrastructure}{HTML}{000000}
\definecolor{colorEnvironment}{HTML}{7570b3}

\newlength\myheight
\newlength\mydepth
\settototalheight\myheight{Xygp}
\definecolor{DiverseMagenta}{rgb}{0.75, 0.0, 0.75}
\definecolor{AccentBlue}{rgb}{0.0, 0.5, 1.0}

\definecolor{mintgreen}{RGB}{152, 255, 152}
\makeatletter
\newcommand*\iftodonotes{\if@todonotes@disabled\expandafter\@secondoftwo\else\expandafter\@firstoftwo\fi}  %
\makeatother

\newcommand{\dimension}[5]{
    \begin{tcolorbox}[fonttitle=\bfseries,coltitle=black,colbacktitle=#2!40,colback=#2!20,colframe=#2,title=#1, after skip=0.35em,left=3pt, right=3pt, top=3pt, bottom=3pt,boxsep=0pt,middle=0.4em,toptitle=6pt, bottomtitle=4pt,sharp corners=all,boxrule=0mm, leftrule=1mm]
    Description: #3
    \tcblower
    Codes: \emph{#4}
    \end{tcolorbox}%
    \noindent#5
}

\newcommand{\underlineColor}[2]{%
  \begingroup
    \setul{0.5ex}{0.3ex}%
    \setulcolor{#1}%
    \ul{#2}%
  \endgroup
}

\begin{document}

\title{A Design Space for Intelligent Agents in Mixed-Initiative Visual Analytics}

\author{Tobias Stähle}
\email{tobias.staehle@inf.ethz.ch}
\orcid{0009-0001-5983-8807}
\affiliation{%
  \institution{ETH Zürich}
  \city{Zürich}
  \country{Switzerland}
}

\author{Matthijs Jansen op de Haar}
\email{mjanse@student.ethz.ch}
\orcid{0009-0005-1673-592X}
\affiliation{%
  \institution{ETH Zürich}
  \city{Zürich}
  \country{Switzerland}
}

\author{Sophia Boyer}
\email{sophia.boyer@inf.ethz.ch}
\orcid{0009-0000-9913-5412}
\affiliation{%
  \institution{ETH Zürich}
  \city{Zürich}
  \country{Switzerland}
}

\author{Rita Sevastjanova}
\email{rita.sevastjanova@inf.ethz.ch}
\orcid{0000-0002-2629-9579}

\affiliation{%
  \institution{ETH Zürich}
  \city{Zürich}
  \country{Switzerland}
}

\author{Arpit Narechania}
\email{arpit@ust.hk}
\orcid{0000-0001-6980-3686}
\affiliation{%
  \institution{The Hong Kong University of Science and Technology}
  \country{Hong Kong SAR}
}

\author{Mennatallah El-Assady}
\email{menna.elassady@inf.ethz.ch}
\orcid{0000-0001-8526-2613}
\affiliation{%
  \institution{ETH Zürich}
  \city{Zürich}
  \country{Switzerland}
}


\renewcommand{\shortauthors}{Stähle et al.}

\begin{abstract}
    Mixed-initiative visual analytics (VA) systems, where human and artificial intelligence (AI) agents collaborate as equal partners during analysis, represented a paradigm shift in human-computer interaction. With recent advances in AI, these systems have seen an increase in sophisticated software agents that have improved task planning, reasoning, and completion capabilities. However, while existing work characterizes agent interplay and communication strategies, there is a limited understanding of the overarching design principles for intelligent agents. Through a systematic review of 90 systems (and 207 unique agents), we propose a design space of intelligent agents comprising six dimensions that collectively characterize an agent's perception, environmental understanding, action capability, and communication strategies. We contribute a novel framework for researchers and designers to explore various design choices for new systems and to situate a system in the current landscape. We conclude with future research opportunities for intelligent agents in mixed-initiative VA systems.
   
\end{abstract}


\begin{teaserfigure}
  \includegraphics[width=\textwidth]{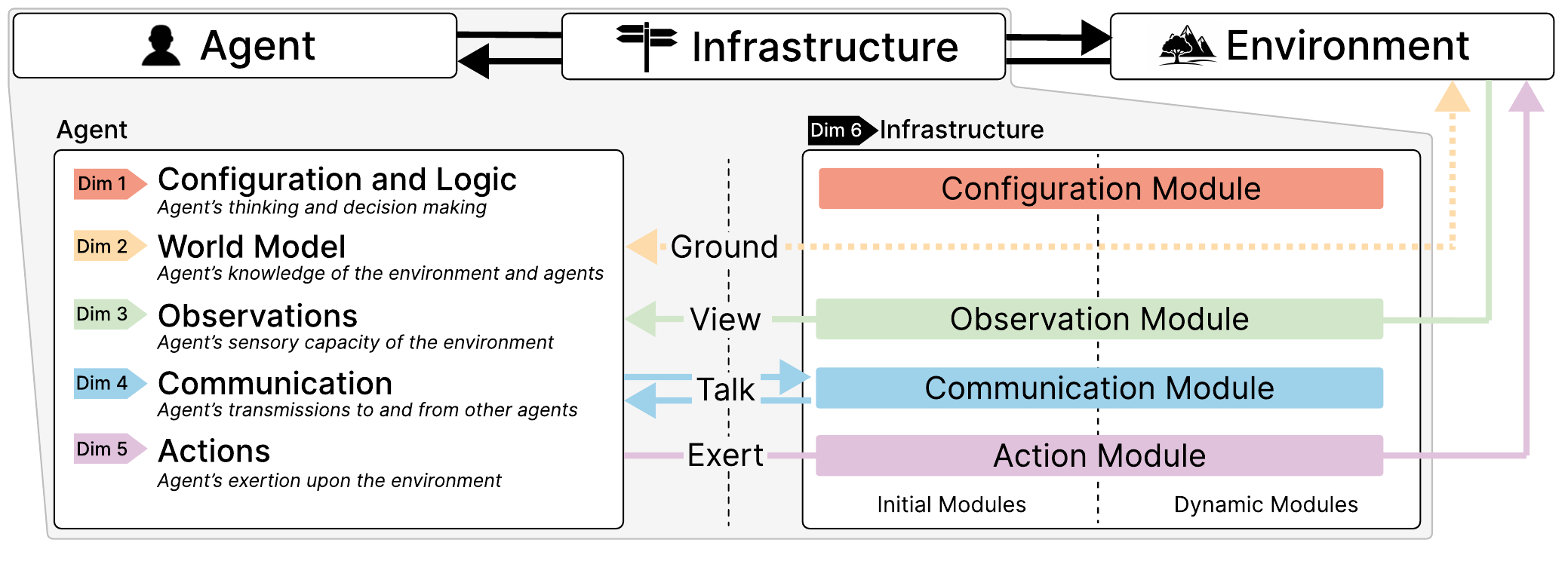}
  \caption{
  To investigate the design space of intelligent agents, a mixed-initiative visual analytics system is partitioned into three components: Agent, Infrastructure, and Environment. Focusing on the agent, the design space characterizes how agents interact with the environment and other agents, given the five dimensions of the agent (Dim 1--5) -- Configuration and Logic, World Model, Observations, Communication, and Actions -- and the description of the Infrastructure (Dim 6).
  }
  \Description{The figure contains three components; Agent, Infrastructure and Environment. The agent and Infrastructure are detailed further with the dimensions in the design space. The first five dimensions belong to the agent: Configuration, World Model, Observations, Communication, and Actions. The sixth dimension is that of the infrastructure, which contains the Configuration Model, Observation Module, Communication Module, and Action Module. Each of these is then split into Initial- and Dynamic Module variants. Finally, multiple arrows indicate the relation between the agent dimensions and the environment; the World Model is grounded within the environment, the Observations view the environment, the Actions exerted onto the environment, and the Communication has multiple agents talking with each other.}
  \label{fig:teaser}
\end{teaserfigure}


\maketitle


\section{Introduction} 

Human-Computer Interaction (HCI) research has long explored ways to optimize collaborations between humans and computers to accomplish tasks \cite{Xu_HAITeaming}, make decisions \cite{Schemmer_HAI_DecisionMaking, reverberi_experimentalHAICollab_2022, LeeHAICollabDecisionMaking2021}, and tackle complex problems effectively \cite{dellermann_hybridIntelligence_2019}. The concept of mixed-initiative interfaces \cite{Horvitz99} has significantly advanced this domain, providing frameworks for collaboration across various fields such as information retrieval \cite{Mass_MIConversationalSearch_2022}, human-computer interaction \cite{Horvitz99}, planning \cite{Chrpa_MIPlanning_2017, Cox_MIPlanningManipulation_2005}, decision-making \cite{Jiang_MIDecisionMaking_HRTeams_2018, Ju_MIDecisionMaking_2022}, and design \cite{Alvarez_FosteringCreativity_2018, Alvarez_StoryDesigner_2022}. In particular, mixed-initiative systems are broadly used in Visual Analytics (VA) applications (e.g., \cite{2018_Podium_EWall,2021_LearningContextualizedUserPreferences_FSperrle,2025_Dango_WHChen, 2015_MixedInitiativeVAUsingTaskDrivenRecommendations_KCook,2025_Leva_YZhao,2025_LightVA_YZhao}), where humans are supported by software in navigating complex analytical tasks.

Amplified by recent advancements, Artificial Intelligence (AI) models are increasingly transforming systems from passive instruments that perform computational operations into intelligent agents that interact within an environment by making observations and taking actions. 
In this context, humans are regarded as agents of a different nature, who employ their capabilities and jointly interact with the environment to carry out analytical tasks \cite{Monadjemi2023}. 
As such, both human and AI agents are considered actors within mixed-initiative (VA) systems  \cite{2024_VisStoryMaker_DJackFreireBraga,2024_JailbreakLens_YFeng, 2020_AnchorViz_JSuh, 2020_VisualInteractionWithDLModelsThroughCollaborative, 2019_VASystemForSubjectMatter_CHagerman, 2023_DASSGood_AWentzel}.
However, the implementation and operational dynamics of these agents vary widely, as do the methods of modeling and employing them. This variation, magnified by inconsistent terminology and varying levels of detail in existing literature and frameworks \cite{Monadjemi2023, Dhanoa_AgenticVisualization_2025}, poses challenges to analyzing, comparing, and evaluating the agents effectively.

Addressing these limitations, in this paper, we propose a novel \textbf{\textit{design space for describing intelligent agents in mixed-initiative VA systems}}. 
In our design space, we consider each system to be comprised of \includegraphics[height=1em]{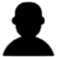}~\textbf{agent(s)}, \includegraphics[height=1em]{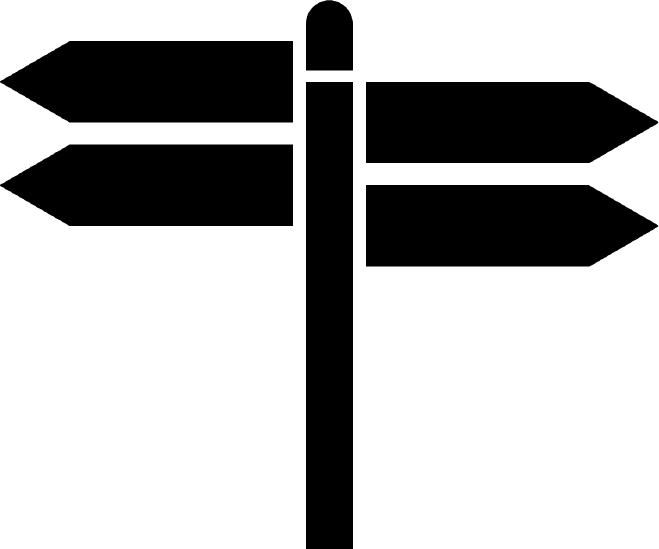}~\textbf{an infrastructure layer}, and \includegraphics[height=1em]{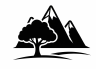}~\textbf{an environment}.
For the systematic categorization, description, and comparison of systems in literature, we further subdivide each system into six dimensions. 
Five dimensions directly pertain to the agent(s): \textbf{(1) \underlineColor{colorConfigurationLogic}{\textit{Configuration + Logic}}}, \textbf{(2) \underlineColor{colorWorldModel}{\textit{World Model}}}, \textbf{(3) \underlineColor{colorObservation}{\textit{Observations}}}, \textbf{(4) \underlineColor{colorCommunication}{\textit{Communication}}}, and \textbf{(5) \underlineColor{colorActions}{\textit{Actions}}}. The sixth dimension relates to the \textbf{(6) \underlineColor{colorInfrastructure}{\textit{Infrastructure}}} that facilitates agent interactions with the environment. 
These dimensions collectively characterize an agent’s perception, environmental understanding, action capabilities,
and communication strategies. We consider the environment to be variable and application dependent, and we do not systematically review it in more detail. 
\autoref{fig:teaser} illustrates the design space, highlighting the various dimensions.

The methodology for developing the design space is rooted in an initial expert discussion, followed by iterative refinement through systematic literature reviews and a detailed examination of specific dimensions. 
Through an analysis of 90 mixed-initiative VA systems and the classification of 207 incorporated agents, we validate the proposed dimensions, thereby offering insights into both existing system features and aspects that remain under-explored.
For instance, we show that in the current systems, the software agents have low autonomy, and the internal adaptation is intrinsic to human agents but only appears sporadically in software agents. 
Also, while it is common to have multiple software agents within a system, only two of the reviewed papers feature multi-stakeholder systems \cite{2012_VAAgentBasedFrameworkForIntrutionAlertAnalysis_RShittu, 2022_HowDoAlgorithmicFairness_CRares}.

To explore potential opportunities beyond implemented and published approaches, by introducing a fictional system, called \textsc{TreeHouse}, we present concrete use cases that illustrate the potential benefits of addressing current gaps, thereby enhancing future system designs. 
For instance, we propose enhancing agent awareness of other agents to improve their interplay and task execution.  
We further emphasize that systems could benefit from incorporating mechanisms for sharing agent availability information, for example, by allowing a human agent to convey their availability to a software agent, in order to reduce their cognitive load.

In summary, our work contributes a unified design space that describes and reconciles the design diversity of intelligent agents in mixed-initiative VA systems. Beyond our survey contribution, we distill current best practices and research gaps, thus encouraging a broader discussion of future opportunities. An interactive overview of our design space, along with the categorization results of the reviewed systems and individual agents, is available at: \href{https://dsagentmi.github.io/?p=AgentDSCHI26}{\texttt{https://dsagentmi.github.io/?p=AgentDSCHI26}}.


\section{Background and Related Work}
In this section, we review foundational works on intelligent agents in mixed-initiative VA systems. Additionally, we provide background information on the frameworks that underpin our design space.
\subsection{Mixed-Initiative in VA Systems}
The concept of human and software agents jointly contributing to tasks, i.e., mixed-initiative interactions, is deeply rooted in HCI. Horvitz's work \cite{Horvitz99} laid out initial principles for designing mixed-initiative interfaces, emphasizing bidirectional collaboration between the user (human) and the (software) agent.
Work by  \citet{Cila22_DesigningHAICollab} explores how to design the behaviors of AI agents in collaboration with human agents by adapting Bratman’s Shared Cooperative Activity framework \cite{Bratman1992}. 
The work emphasizes that software agents must clarify and clearly communicate their intent and be prepared to both offer and request help.  
More general design guidelines on how human-AI interaction should be implemented are proposed and evaluated by  \citet{Amershi2019}.

The concept of '\textit{mixed-initiative}' has been introduced into the field of VA by several works, such as  \citet{2006_InteractiveVisualSynthesisOfAnalyticalKnowledge_DGotz},  \citet{2007_ContextAwareAdaptiveInfromationRetrieval_ZWen},  \citet{2007_Tibor_DLiu}, and  \citet{2015_MixedInitiativeVAUsingTaskDrivenRecommendations_KCook}.
Later,  \citet{Buehler2025} introduced the concept "AI-in-the-loop", which suggests integrating AI as a collaborative partner in VA applications.
Although mixed-initiative systems have been developed for a wide range of VA application scenarios (see, e.g., \cite{2018_Podium_EWall, 2019_VAForTopicModeling_MElAssady, 2015_MixedInitiativeVAUsingTaskDrivenRecommendations_KCook,2016_Voyager_KWongsuphasawat}), they are especially known in the context of user guidance, which explores how automated guidance can complement user-driven analysis. For example,  \citet{Makonin2016} define a conceptual framework for a mixed-initiative VA system that comprises five components: data wrangling, alternative discovery and comparison, parametric interaction, history tracking and exploration, and system agency and adaptation, to enable seamless collaboration between human reasoning and machine prediction.
Other efforts have framed guidance as an explicit task taxonomy \cite{Ceneda2019, PerezMessina2022} and emphasize this form of software agent assistance.  \citet{Sperrle2020} introduce the concept of co-adaptive guidance, which describes the process of aligning the human and software agents' understanding of the task and analysis process, and characterizes learning and teaching, i.e., guidance interactions that occur in mixed-initiative VA.
\subsection{Human and Software as Intelligent Agents}
A recent line of work explicitly treats both human analysts and models/algorithms as agents within the VA process, and several theoretical frameworks have been proposed to characterize them. 
For instance, \citet{Monadjemi2023} provide an agent-based conceptual model for mixed-initiative VA by drawing from the AI agent paradigm. 
Their work characterizes the analytical process through a multi-agent perspective, while distinguishing between human and artificial agents working in a joint environment. 
Agents observe the VA environment and take actions, thus unifying traditional VA, guided analytics, and collaborative analysis under the common language of agents. 
Therefore, human analysts, automated guidance, and analytic settings can be characterized in parallel and are modeled as interacting agents. 
 \citet{Holter2024} observe that this agent-centric view is well-suited to capture the complexities of mixed-initiative VA. 
They note that using an agent-based lens to study VA has proven effective for more complex mixed-initiative and guided collaborative systems. 
Furthermore, they identify three main dimensions for Human-AI collaboration: agency \cite{Monadjemi2023, El-AssadyM22_Bias_Reasoning_Pitfalls}, i.e., which agent is in control of the task-solving process, interaction \cite{Amershi2019, El-AssadyM22_Bias_Reasoning_Pitfalls, 2021_LearningContextualizedUserPreferences_FSperrle}, describing how the agents collaborate and communicate, and adaptation \cite{2021_LearningContextualizedUserPreferences_FSperrle, WhereAreWeSoFar_Li_2024}, i.e., which agent adapts over time by observing, interacting and receiving feedback from other agents and its environment.
More recently,  \citet{Bernard2025} introduced the Human-Data-Model Canvas that characterizes the roles of human agents, data, and software agents. This work proposes a method to characterize how human and software agents benefit and contribute within the VA process. Furthermore,  \citet{Dhanoa_AgenticVisualization_2025} propose a collection of design patterns with respect to agentic roles drawn from general data science roles \cite{DataScienceRoles_Crisan_2021}, communication, and coordination in agentic visualizations. 
Although several theoretical frameworks exist that describe human and software agents, they abstract away several important agent characteristics. These frameworks treat each participant (human or software/tool) as a generic agent without spelling out the agent's internal decision logic, world model -- abstract interpretation of the environment, or communication mechanism. 
This limits the ability to describe agents in mixed-initiative VA systems accurately, weakens the discussion and comparison between human and software agents, and hampers cross-system comparisons within the research community.
In our design space, we bridge high-level HCI design taxonomies with the descriptive model of intelligent agents found in mixed-initiative VA systems. By detailing how human and software agents perceive, reason about, and act on the analytic environment, we offer a novel design space that complements and extends prior frameworks with a common language for thinking about and designing the agents in mixed-initiative VA.


\section{Methodology and Definitions}
\label{sec:methodology}
\begin{figure}[H]
    \centering
    \includegraphics[width=1\linewidth]{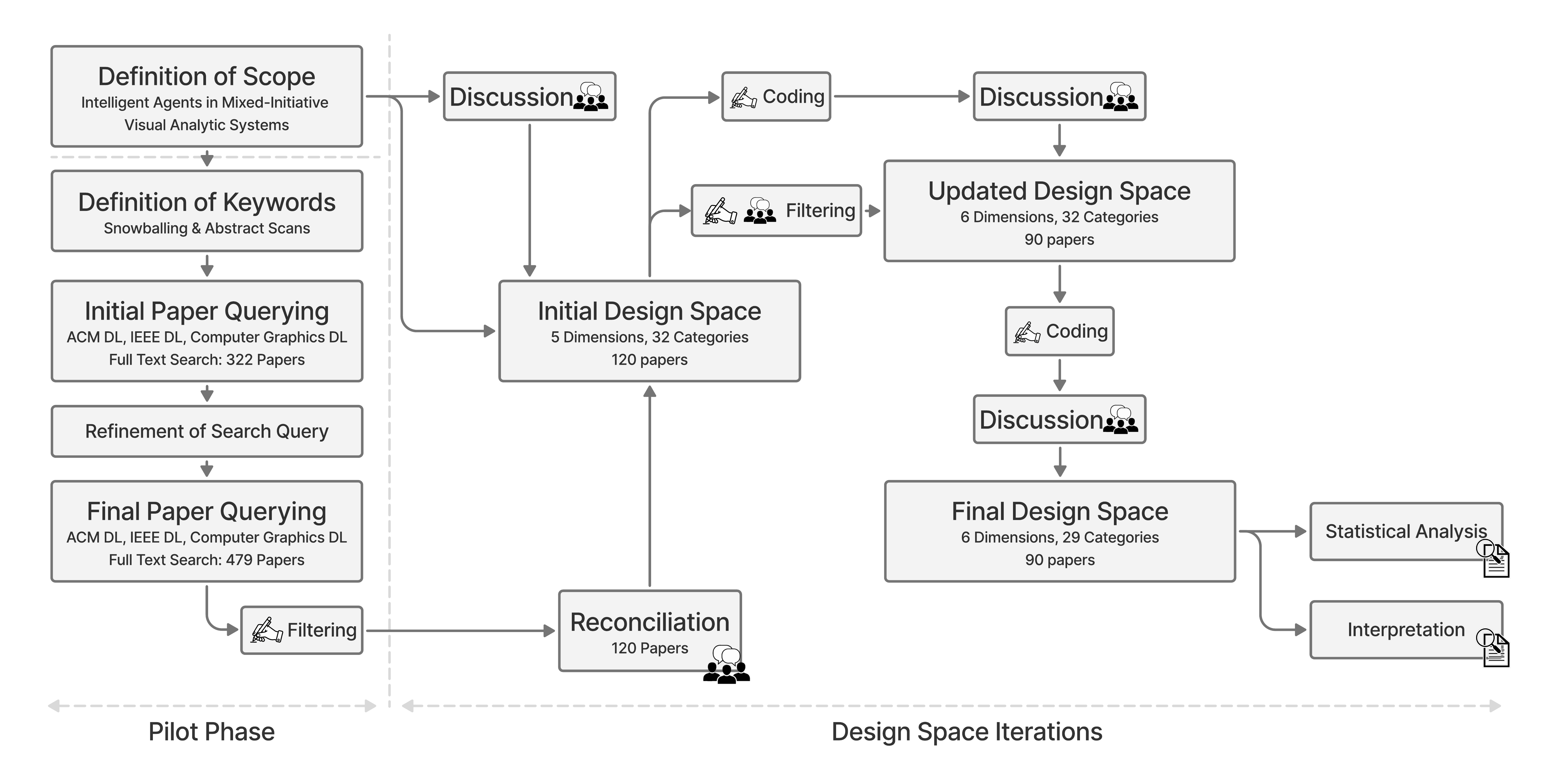}
    
    \caption{Methodology: The design space creation consists of three phases -- definition of scope, pilot phase, and design space iterations. First, the scope is defined to be able to identify intelligent agents in a mixed-initiative VA system. Three of the authors review and filter existing works. Then we construct the design space based on initial discussions among all authors and three iterations of paper coding, additional discussions among the authors, and design space refinements. }
    \Description{The figure describes the general methodology in the form of a workflow. Three main areas are described, namely: the definition of scope, the pilot phase, and the design space iterations. The definition of scope is the initial step of the workflow. The pilot phase shows a five-phase workflow, which then leads into the design space iterations. These include: (1) Define Keywords, into (2) Initial Paper Querying, into (3) Refine Search Query, into (4) Final Paper Querying, into (5) Filtering. Three points then lead into the design space iterations, these being: Discussion, Definition of Scope, and Filtering. With these three, an initial design space was drafted. After which, two consecutive workflows were initiated. The first being coding, and then discussion. The second is just filtering. Both of which led to an Updated Design Space. Afterwards, a final workflow is initiated, consisting of three phases. Namely: (1) Coding, into (2) Discussion, into (3) Final Design Space. With the Final Design Space, both a Statistical Analysis and an Interpretation were conducted.}
    \label{fig:methodology}
\end{figure}

The creation of our design space is grounded in the definition of the scope, a systematic literature review, and collaborative design space iterations that are outlined in \autoref{fig:methodology}. Our process is based on an open coding methodology that is used for systematic analysis of topics and trends in the field of HCI research \cite{BaeZWDHS22, Lee2024, Sperrle2020, Chan2025}. 

\subsection{Scope and Definitions} 
\label{sec:ScopeAndDefinitions}
In the absence of an established unified definition for an intelligent agent, and acknowledging the challenge of precisely defining the terms 'agent' and 'intelligence' \cite{Wooldridge_Jennings_1995}, we first provide several perspectives of other works on this concept.  

In the domain of Reinforcement Learning, \citet{Abel2023ContinualReinforcementLearning} state that the term "agent" frequently refers to a collection of mathematical functions that delineate the behavior of the agent. Moreover, they note that agents are capable of modifying their internal parameters, enabling them to learn and adapt in response to past observations and actions.
Research in the broader field of AI by  \citet{RusselNorvig1995} defines an agent as an entity ``\emph{that receive percepts from the environment and
perform actions}.'' In their understanding, the internal unit of the agent is mapping percepts to specific actions, enabling the agent to act upon different scenarios. 
 \citet{rousseau1997mixed} extend this concept by defining software agents as computer systems that utilize their own knowledge bases, possess distinct goals and capabilities, execute actions, and engage with other agents as well as with humans. Additionally, they emphasize autonomy as a critical attribute of such agents, demonstrated when an agent independently initiates actions.
 
Moreover, research conducted by  \citet{Wooldridge_Jennings_1995}, as well as  \citet{Weiss1995MultiAgent}  characterizes intelligent agents as entities that exhibit specific behaviors to achieve their assigned objectives. These behaviors include proactiveness, which involves taking initiative; reactivity, which entails responding promptly to changes; and social ability, which refers to interacting effectively with other agents.
In addition, current works by  \citet{shavit2023practices} and \citet{Chan2023AgenticAlgorithms} manifest that to be considered an agent, the entity must have an active and measurable environmental impact. It should be capable of taking actions that meaningfully and persistently alter its environment. These changes should be observable and/or measurable by the agent itself or external observers. Simple input-output mappings do not qualify as agentic under this requirement.

In the following, we summarize the existing definitions of mixed-initiative systems and visual analytics.
\begin{itemize}
   
     \item[] \textbf{\textsc{Mixed-initiative Systems: }}  \texttt{Mixed-initiative refers to a flexible interaction strategy, where each agent can contribute to the task what it does best.} --\citet{Horvitz99}
     \item[] \textbf{\textsc{Visual Analytics: }}  \texttt{ Visual analytics combines automated analysis techniques with interactive visualizations for an effective understanding, reasoning, and decision-making on the basis of very large and complex data sets.} --\citet{Keim_VA_2008}
\end{itemize}

In addition, we consolidate the different definitions of an intelligent agent to capture their most important aspects.

\begin{itemize}
    \item[] \textbf{\textsc{Intelligent Agent:} } \texttt{In mixed-initiative (VA) systems, an intelligent agent is any entity, human or computational, capable of perceiving its environment and proactively performing goal-driven actions.  An agent must be capable of \underline{autonomous decision-making}}
    \footnote{The agent crucially must take \textbf{autonomous decisions and actions} that significantly influence system behavior beyond purely algorithmic calculations.}
    \texttt{by interpreting its environment and applying internal logic, allowing it to take or cede initiative.} 
\end{itemize}

These definitions scope and define the inclusion criteria for the works considered in the literature search and are consistent with the objectives of our work.

In summary, the scope of our research encompasses systems in which at least one software agent and one human agent coexist and can interact with each other, aiming to evaluate data using visual processing and gain valuable insights. Furthermore, in the appendix \autoref{appendix:exampleExclusion}, we provide an example of work that was excluded from our discussions and an explanation of the respective exclusion criteria.
 
\subsection{Pilot Phase}
Based on the defined scope, we conducted a thorough literature review of existing mixed-initiative VA systems. Important resources here are publications in the fields of HCI, data visualization, and VA. For this purpose, we searched the ACM Digital Library, the IEEE Digital Library, and the Computer Graphics Digital Library. This way, we ensured the inclusion of all publications from leading conferences and journals, such as CHI, IUI, VIS, VAST, UIST, TVCG, EuroVis, and EUROGRAPHICS within our search scope. We only consider complete research articles or conference reports that describe an implemented system.

\paragraph{Search Query} We identified relevant keywords from a set of anchor papers using titles, author keywords, and abstracts. In the initial search, we searched for (“agent” OR “assist*”) AND (“mixed-initiative”) AND (“visual analytics” OR "visual analysis"). This search query (performed in July 2025) resulted in  322 papers. After further discussion among all authors, the search scope was adapted by removing the search terms “agent” OR “assist*“, deliberately creating a comprehensive selection of relevant works that include any variations or synonyms for agents and assistants. The final search query is (“mixed-initiative”) AND (“visual analytics” OR "visual analysis"). Both mixed-initiative and VA are established terms within the community and thus present an exhaustive representation of the scope. This specific search query yielded 479 articles for the queried digital libraries in July 2025.
\paragraph{Article Filtering} Based on the discussion of the original scope among all authors, three authors jointly filtered all articles found by the search query to verify whether they fit the defined scope. The initial filtering was performed to identify all articles that fell within the broader scope of the specified topic. All authors independently performed the filtering to prevent biases. To maintain the consistency of the articles considered, regular meetings were held to discuss ambiguous cases and refine the scope as necessary. 
Initially, a subset of articles was used, which was filtered by all three authors and then discussed to achieve agreement.
Then, each article was evaluated by at least two annotators. A satisfactory agreement between the annotators of 95\% was achieved. During the joint discussions, disagreements were discussed and, if necessary, resolved by the main author as he has reviewed the entire corpus. The filtering process resulted in a corpus of 120 papers. The main exclusion criteria are: The paper should clearly describe a mixed-initiative system and include at least one software agent and one human agent. For exhaustiveness, Wizard-of-Oz studies were included; however, robotic agents were excluded due to their physical nature.
Furthermore, the described system should be within the VA domain, and the software agent needs to take a prominent role within the system, contributing to at least one analytic task type, similar to the analytic task types identified by Sperrle et al. \cite{Sperrle_SurveyHCEvaluationHCML_2021}. Additionally, the software agents need to execute tasks that are more than one-step calculations (e.g., data projections, dimensionality reduction). Papers that did not include examples of a system, like frameworks, surveys, design spaces, taxonomies, and typologies, were excluded. 

\subsection{Design Space Creation \& Iteration} 
 After filtering, the authors collaboratively developed an initial design space draft through meetings, utilizing their experience, discussions, and several trial iterations to code intelligent agents across multiple dimensions. Each dimension consists of several categories that hold codes -- category values -- by which the individual agents differ. We did not strictly enforce the codes to be mutually exclusive, as multiple codes applied to some agents in the examined systems. In addition, an agent could hypothetically assume any permutation of codes. When coding the individual agents, patterns emerge from the trends in the design of the agents that are based on a common code combination in the design space.
\paragraph{Design Space Iterations} The initial design space consisted of five dimensions and 32 categories. Using a mixed-methods approach, the design space was iteratively refined in subsequent coding iterations. To this end, each article contained in the corpus was analyzed by at least two authors, and each agent identified in the respective system was coded based on the most recent iteration of the design space. In regular meetings, the coding was discussed batch-wise, and, if necessary, the design space was updated to add new interesting categories and codes or to combine redundant ones. Based on the discussions, the authors recalibrated and restarted the tagging process. Furthermore, a more detailed reading of the corpus continuously refined the scope, resulting in a joint case-by-case evaluation in which another 30 articles were excluded from the initial corpus by the authors. This resulted in 90 articles that were contained in the final scope.
This iterative process was repeated three times until all authors agreed on the design space and all identified agents had been tagged. To ensure consistency, each article was coded again by at least two annotators. Disagreements per agent within each paper were counted and weighted relative to the number of agents in the system and the number of dimensions in the design space. 
This resulted in a 93.5\% (std: 0.04) agreement rate. Any disagreements were discussed and, if necessary, resolved by the first author.
After all three iterations, the final design space now consists of six dimensions and 29 categories.

\section{Design Space for Intelligent Agents in Mixed-Initiative VA} 

Our design space builds on the idea of modeling humans and software agents in the same way while externalizing the agents that exist in the mixed-initiative VA system, similar to Monadjemi et al. \cite{Monadjemi2023}. This allows us to characterize the agents separately from the actual application and identify similarities and differences when comparing agents from one system or agents from different systems.
To this end, we divide the mixed-initiative VA system under consideration into three parts, as shown in \autoref{fig:teaser}: \textit{Agent(s)}, \textit{Infrastructure Layer}, and \textit{Environment}. In the following, we first characterize each individual part.

\paragraph{ \includegraphics[height=1em]{figures/AgentIcon.pdf}\textbf{Agent(s)}} We consider humans and software agents as defined in \autoref{sec:ScopeAndDefinitions} as agents. An intelligent agent is a self-contained entity that can perceive its environment and interact with it through its capabilities. "Intelligent" refers to the ability to think and make decisions to a certain extent by incorporating its internal, abstract interpretation of its environment – the \textit{World Model}. 
This design space focuses on the agent, which is why most dimensions of the design space are focused on characterizing the agent more precisely. Specifically, an agent is described through the dimensions Configuration and Logic, World Model, Observations, Communication, and Actions, which are explored in the following subsections. 
For agents to interact with the environment, an additional infrastructure layer is required that provides the necessary default settings and functionalities through various modules. This allows the system developer to integrate default settings, restrictions, rules, and additional dynamics into the mixed-initiative VA system without having to define them for each agent.

\paragraph{\includegraphics[height=1em]{figures/InfrastructureIcon.pdf}\textbf{Infrastructure Layer}} The infrastructure layer enables the successful interaction of the agents with each other and with the environment. There are various modules for this, which can be initially configured or dynamically adjusted using logic during the run-time of the application. The \textit{Configuration Module} contains basic parameters for designing the individual agents. The \textit{Observation Module} and the \textit{Action Module} are each designed so that the flow of information can only occur in one direction. The \textit{Observation Module}  provides the agent with information exclusively from the environment. The \textit{Action Module} allows the agent to perform actions in the environment. Both modules also define which areas and data the agent has access to in the environment for observation and which actions are available to the agent for execution. The \textit{Communication Module} describes the possible ways and types of communication between agents. Possible ways of communication here refer, e.g., to which agents can communicate with and in which format, such as single channel, group channel, or broadcast. It also provides the communication channel, since in this design space, communication between agents takes place outside of the environment.
The \textit{Infrastructure Layer} is an integral part of a mixed-initiative VA system, which includes intelligent agents. When observed through the lens of this design space, it serves as the portal for agents to interact with the VA environment. This also makes it an important part when characterizing the agents of such a system. Hence, our design space dedicates a dimension to the infrastructure layer.

\begin{figure}[H]
    \centering
    \includegraphics[width=\textwidth]{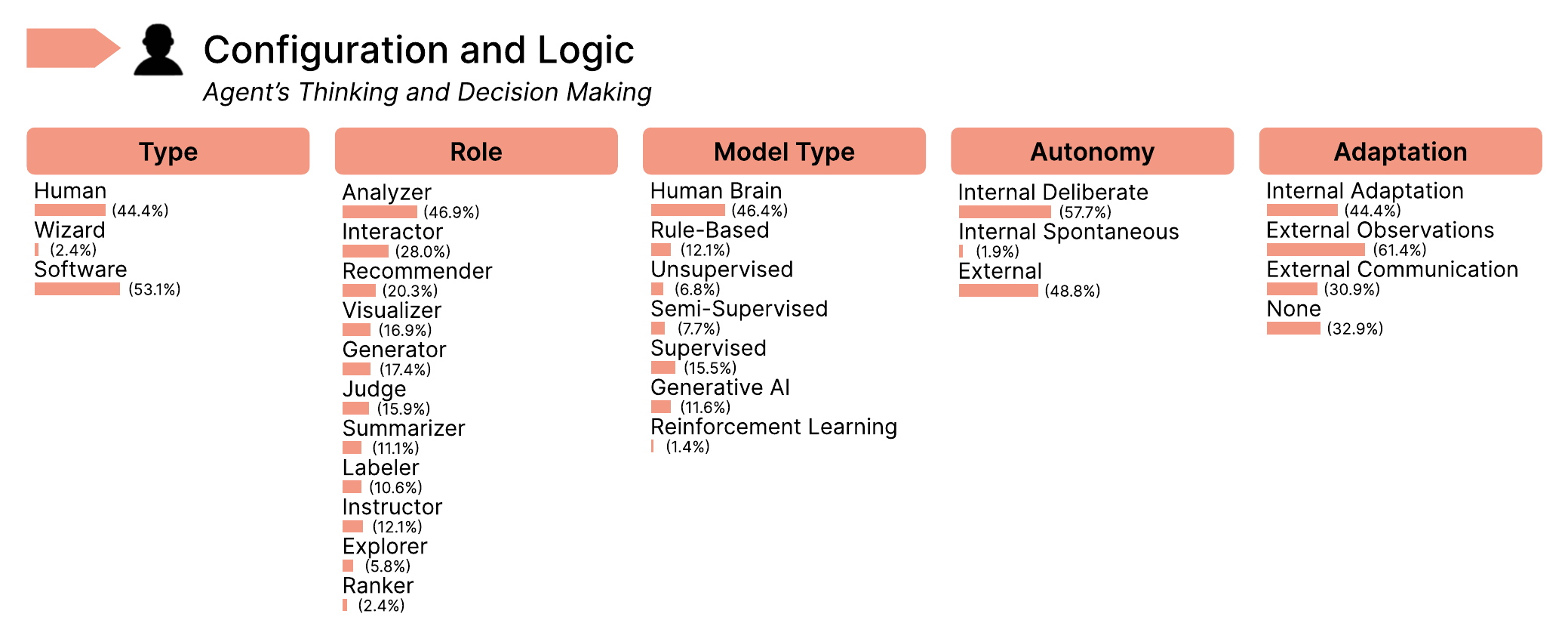}
    \caption{Configuration and Logic: categories, codes, and the percentage frequency of each code's occurrence in the investigated agents}
    \Description{The figure shows all of the categories within the configuration and logic dimension, namely: Type, Role, Model Type, Autonomy, and Adaptation. Within each category, all codes and the frequency (as a percentage) relative to the number of agents are also shown.}
    \label{fig:codingConfigurationLogic}
\end{figure}

\begin{figure}[H]
    \centering
    \includegraphics[width=\textwidth]{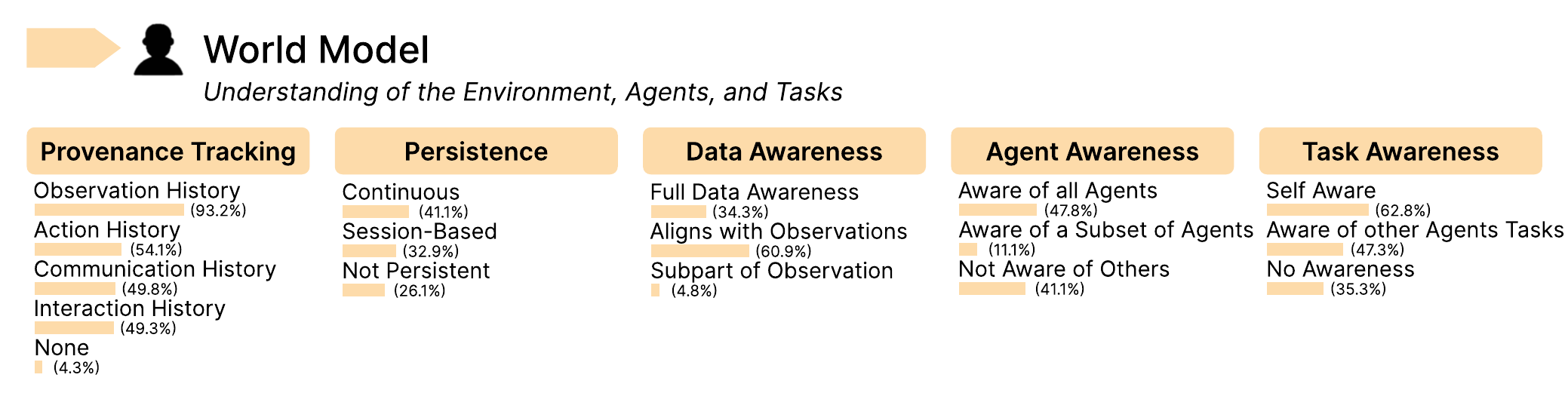}
    \caption{World Model: categories, codes, and the percentage frequency of each code's occurrence in the investigated agents}
    \label{fig:codingWorldModel}
    \Description{The figure shows all of the categories within the world model dimension, namely: Provenance Tracking, Persistence, Agent Awareness, Data Awareness, and Task Awareness. Within each category, all codes and the frequency (as a percentage) relative to the number of agents are shown as well.}
\end{figure}

\begin{figure}[H]
    \centering
    \includegraphics[width=\textwidth]{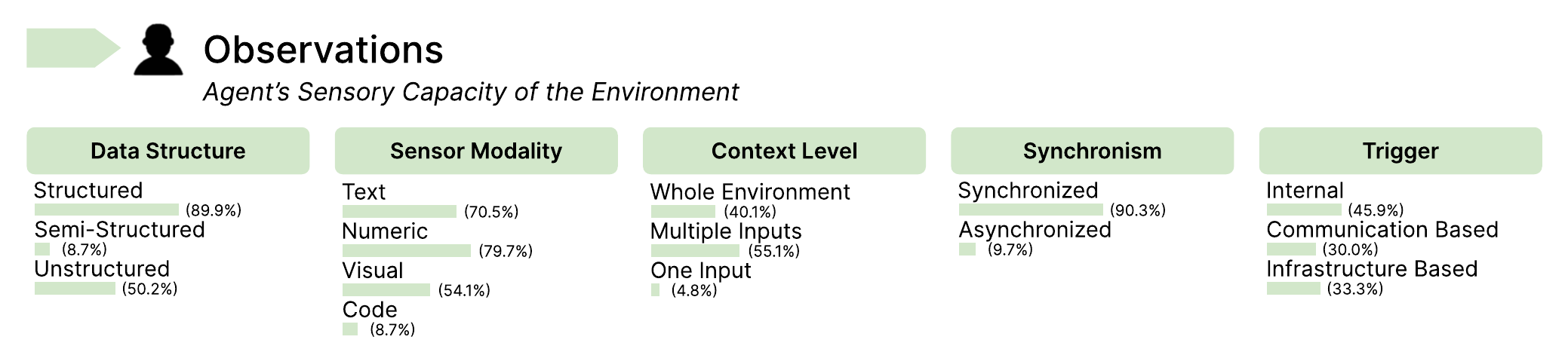}
    \caption{Observations: categories, codes, and the percentage frequency of each code's occurrence in the investigated agents}
    \label{fig:codingobservation}
    \Description{The figure shows all of the categories within the observations dimension, namely: Data Format, Sensor Modality, Synchronism, Context Level, and Trigger. Within each category, all codes and the frequency (as a percentage) relative to the number of agents are shown as well.}
\end{figure}

\begin{figure}[H]
    \centering
    \includegraphics[width=\textwidth]{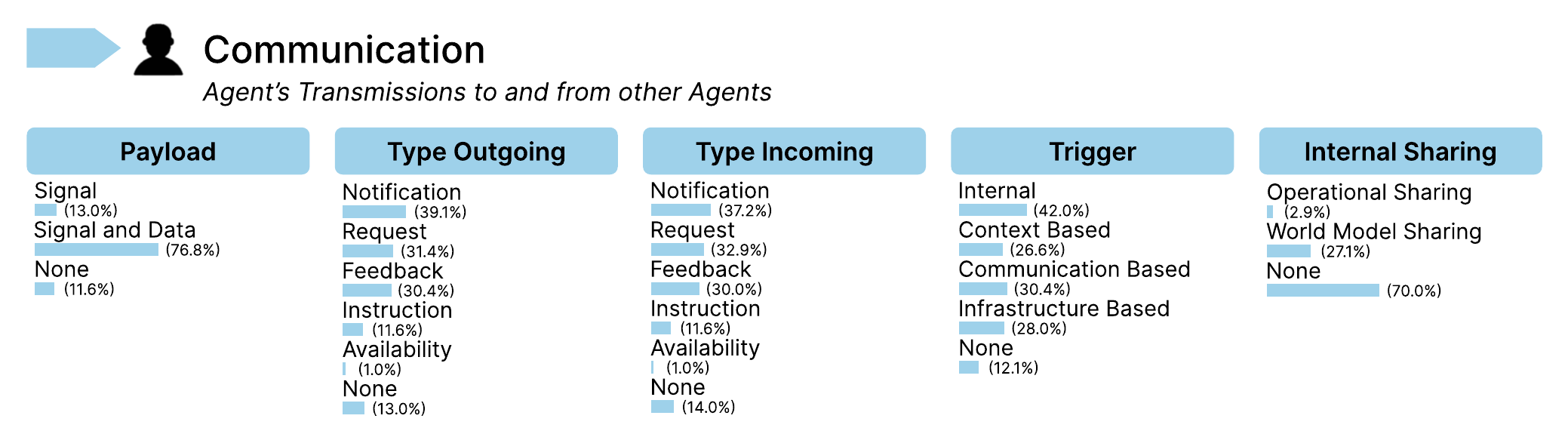}
    \caption{Communication: categories, codes, and the percentage frequency of each code's occurrence in the investigated agents}
    \Description{The figure shows all of the categories within the communication dimension, namely: Payload, Type Outgoing, Type Incoming, Status Sharing, and Trigger. Within each category, all codes and the frequency (as a percentage) relative to the number of agents are shown as well.}
    \label{fig:codingCommunication}
\end{figure}

\begin{figure}[H]
    \centering
    \includegraphics[width=\textwidth]{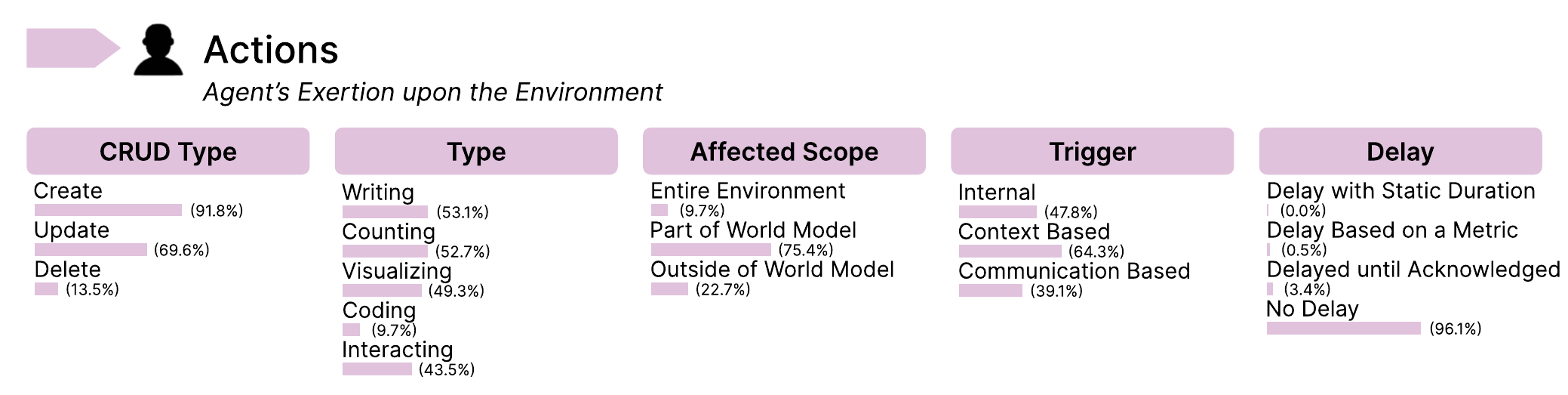}
    \caption{Actions: categories, codes, and the percentage frequency of each code's occurrence in the investigated agents}
    \label{fig:codingactions}
    \Description{The figure shows all of the categories within the actions dimensions, namely: CRUD Type, Type, Action Scope, Trigger, and Delay. Within each category, all codes and the frequency (as a percentage) relative to the number of agents are shown as well.}
\end{figure}

\begin{figure}[H]
    \centering
    \includegraphics[width=\textwidth]{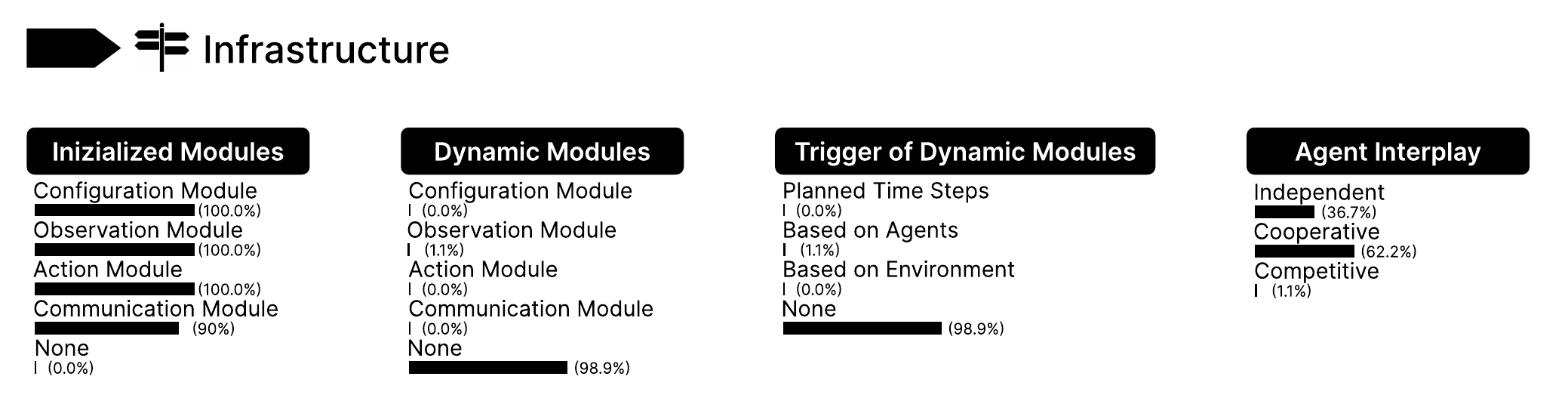}
    \caption{Infrastructure: categories, codes, and the percentage frequency of each code's occurrence in the investigated agents}
    \Description{The figure shows all of the categories within the infrastructure dimension, namely, Initialized Modules, Dynamic Modules, Trigger of Dynamics, and Agent Interplay. Within each category, all codes and the frequency (as a percentage) relative to the number of systems are shown as well.}
    \label{fig:codingInfrastructure}
\end{figure}

\paragraph{\includegraphics[height=1em]{figures/EnvironmentIcon.pdf}\textbf{Environment}} Similar to the understanding in other works \cite{Monadjemi2023, Abel2023ContinualReinforcementLearning}, in our approach, the environment describes the general VA system and all its components, such as data management, analytical models, visualizations, user interface, knowledge representation, and the possibility of evaluation and feedback loops. As described above, the infrastructure layer is necessary for agents to be able to interact with the environment. This ensures that the functionalities and access of the agents to parts of the environment are clearly defined. The environment reflects the current state of progress within the VA system at any given time. Inter-agent communication and the agents themselves are not part of the environment. Nevertheless, agents can observe the actions of other agents in the environment, as these actively edit the environment with their actions. While the environment is a substantial part of a mixed-initiative VA system, this design space focuses on the agents themselves; thus, it does not characterize the environment in detail. 
\paragraph{Final Version of the Design Space} Through the iterative process of coding and updating the design space, we reached the final version of our design space, which contains six dimensions – \textit{Configuration + Logic}, \textit{World Model}, \textit{Observations}, \textit{Communication}, \textit{Actions}, and  \textit{Infrastructure Layer}. In total, the dimensions have 29 categories and 117 codes. In total, we examined 90 systems. Since our design space refers in detail to the agent characteristics, we coded each agent in these systems individually, resulting in 207 agents.
In the following, we explain the individual dimensions, subcategories, and codes of our design space to provide a detailed overview of the design space. Note that in the "Detailed Code" figures (\Cref{fig:codingConfigurationLogic,fig:codingWorldModel,fig:codingobservation,fig:codingCommunication,fig:codingactions,fig:codingInfrastructure}), we list the percentiles of all agents that meet the definition of each code. Some agents cannot be mapped exclusively to one code and instead meet multiple code definitions simultaneously. Accordingly, the sum of all codes within a category may be greater than 100\%. In addition,  \autoref{appendix:usecases} provides two compelling use cases that demonstrate the practical application of our design space. Further, an overview with all definitions of the categories and codes is provided in  \autoref{appendix:definitions}.

\subsection{Configuration and Logic \includegraphics[height=1em]{figures/AgentIcon.pdf}}
The Configuration and Logic dimension describes the internal computational unit of agents, i.e., whether this is the human brain or a set of algorithmic instructions. The logic constitutes an agent's internal decision-making process, and more generally, shapes the agent's overall structure. Further, within this dimension, we characterize the overall role the agent takes within the system to reason about tasks the agent covers. Further, we examine the autonomy of the agent and possible adaptation to reason about the agent's behavior. While in the design space, we consider both entities -- human and software -- as agents, we label the agent type within this dimension. However, some codes exclusively apply to either human or software agents, and some attributes are automatically assigned to human agents (e.g., Model Type).

\dimension{Type}{colorConfigurationLogic}{What is the type of agent?
}{Human, Wizard, Software}
                
This category describes the manifested form of an agent. As previously stated, we group all agents within the design space regardless of this manifested form; nevertheless, in this category, we distinguish between \underlineColor{colorConfigurationLogic}{Human} agents and \underlineColor{colorConfigurationLogic}{Software} agents. In each analyzed system, there is at least one human agent, the user. Additionally, there exist studies which describe \underlineColor{colorConfigurationLogic}{Wizard} agents \cite{2022_ConverseWithAnalyticChatbot_VSetlur, 2025_WizzardGuidanceStrategiesAndDynamics_FSperrle,2025_GuidanceSourceMatters_ANarechania, 2024_Socrates_GWu,2019_DoWhatIMean_MTroy}. At the same time, these simulated software agents still retain some human-like qualities and thus qualify as a distinct agent type. In the Model Type category, both human agents and most wizard agents are labeled as Human Brain.

\dimension{Role}{colorConfigurationLogic}{What are the roles that an agent takes within a given system?}{Analyzer, Instructor, Interactor,  
Recommender, Explorer, Visualizer, 
Generator, Ranker, Judge, 
Summarizer, 
Labeler}

We assigned roles held by different types of agents to identify similar agents based on their tasks and responsibilities within a given system. Considering that there are a large amount of tasks present in VA, we distinguish whether the agent is analyzing data (\underlineColor{colorConfigurationLogic}{Analyzer}, e.g., \cite{2012_VAAgentBasedFrameworkForIntrutionAlertAnalysis_RShittu,2019_UserbasedVAWorkflowForEMA_DCashman, 2020_explAIner_TSpinner}), instructs or orchestrates other agents (\underlineColor{colorConfigurationLogic}{Instructor}, e.g.,  \cite{2025_LightVA_YZhao,2016_Eviza_VSetlur}), interacts with the UI (\underlineColor{colorConfigurationLogic}{Interactor}, e.g.,  \cite{2012_InteractiveAnalysisOfBigData_JHeer,2025_WhatIF_AMishra,2018_Podium_EWall}), recommends data, insights or facts (\underlineColor{colorConfigurationLogic}{Recommender}, e.g., \cite{2025_WizzardGuidanceStrategiesAndDynamics_FSperrle,2016_Voyager_KWongsuphasawat,2024_GuidedByAI_SHa, 2019_AugmentingVisualizationsWithInteractiveDataFacts_ASrinivasan}), explores data (\underlineColor{colorConfigurationLogic}{Explorer}, e.g., \cite{2006_InteractiveVisualSynthesisOfAnalyticalKnowledge_DGotz, 2016_TimeFork_SKarthik, 2021_QualitativeCausalModeling_FHusain, 2024_DiscipLink_CZheng}), visualizes data (\underlineColor{colorConfigurationLogic}{Visualizer}, e.g., \cite{2025_WhatIF_AMishra,2024_InkSight_YLin,2022_HowDoAlgorithmicFairness_CRares,2025_Divisi_VSivaraman, 2016_Voyager_KWongsuphasawat}), generates new data which is not yet within the environment (\underlineColor{colorConfigurationLogic}{Generator}, e.g., \cite{2025_Sprout_YLiu,2016_TimeFork_SKarthik,2025_VisCars_PMoens}), ranks datapoints based on internal criteria (\underlineColor{colorConfigurationLogic}{Ranker}, e.g., \cite{2018_Podium_EWall, 2024_CalliopeNet_QChen}), evaluates results and data (\underlineColor{colorConfigurationLogic}{Judge}, e.g.,  \cite{2022_VideoModerator_TTang, 2024_Stile_SKabir,2024_TheDataSaysOtherwise_YFu}), summarizes data, steps taken and results (\underlineColor{colorConfigurationLogic}{Summarizer}, e.g., \cite{2024_Socrates_GWu, 2024_iToT_ABoyle, 2025_Dango_WHChen}), or labels data in a semi-supervised context (\underlineColor{colorConfigurationLogic}{Labeler}, e.g., \cite{2025_KMTLabeler_HWang, 2025_HEPHA_SZhou, 2022_TowardsVisualExplainableActiveLerning_SJia}).

\dimension{Model Type}{colorConfigurationLogic}{What learning techniques are used by an agent for thinking and decision-making?}{Rule-Based, Supervised, Unsupervised, Semi-Supervised, Reinforcement Learning, Generative AI, Human Brain}

It is essential to properly capture an \textit{intelligent agent's} internal decision-making unit, as this impacts the way an agent thinks, acts, and what level of task complexity it can solve. Human and wizard agents are classified by assigning the model type \underlineColor{colorConfigurationLogic}{Human Brain}. Here, we define an array of labels to describe the variety of decision-making units observed in different software agents.
Early mixed-initiative VA systems primarily used \underlineColor{colorConfigurationLogic}{Rule-Based} software agents \cite{2006_InteractiveVisualSynthesisOfAnalyticalKnowledge_DGotz,2010_PredictiveAnalyticsUsingBlackboard_JYu,2011_GuidingFeatureSubsetSelection,2012_InteractiveAnalysisOfBigData_JHeer}. At the same time, the advances in machine learning in recent years have enabled machine learning techniques that we categorize as: \underlineColor{colorConfigurationLogic}{Supervised}, \underlineColor{colorConfigurationLogic}{Unsupervised}, \underlineColor{colorConfigurationLogic}{Semi-Supervised}, or \underlineColor{colorConfigurationLogic}{Reinforcement Learning}. While \underlineColor{colorConfigurationLogic}{Supervised} models are used, e.g., for classifying data \cite{2025_HEPHA_SZhou, 2024_GuidedVAForImageSelectionInTimeAndSpace_IPerezMessina, 2022_IntentVizor_GWu} and learning user preferences \cite{2018_Podium_EWall, 2021_LearningContextualizedUserPreferences_FSperrle}, \underlineColor{colorConfigurationLogic}{Unsupervised} models are applied in agents that, e.g., recommend similar data points \cite{2021_QuestionComb_RSevastjanova,2022_SupportingSerendipitousDiscovery_MJasim, 2022_DeconstructingCategorizationInVisualizationRecommendation_DJung}. In current systems, \underlineColor{colorConfigurationLogic}{Semi-Supervised} learning is used for agents that cover active-learning scenarios \cite{2025_KMTLabeler_HWang,2021_IntegratingPriorKnowledge_APister}  and \underlineColor{colorConfigurationLogic}{Reinforcement Learning}, e.g., for agents that optimize visual space to visualize data insights \cite{2025_InsigHTable_GLi}, or identify and recommend interesting patterns in the data \cite{2024_SupportingGuidedEVAOnTimeSeriesData_YShi}.
In addition, we separate \underlineColor{colorConfigurationLogic}{Generative AI} from the other model types to symbolize the increasing importance that is present in agents for VA systems. Multiple systems use this technology for agents that, e.g., analyze textual data \cite{2023_Datatales_NSultanum, 2024_DiscipLink_CZheng}, generate automation scripts \cite{2025_Dango_WHChen} or visualizations generation \cite{2025_LightVA_YZhao}, and cover analytical tasks in general \cite{2025_LightVA_YZhao, 2025_Leva_YZhao}.

\dimension{Autonomy}{colorConfigurationLogic}{Is the agent able to make decisions and act by itself or does it require external stimuli?}{Internal Deliberate, Internal Spontaneous, External}

According to the \textit{Agency} dimension in the design space presented in "Deconstructing Human-AI Collaboration" by Holter and El-Assady \cite{Holter2024}, within a mixed-initiative system, agency can be distributed among all agents. Therefore, each agent has a certain degree of autonomy, which impacts when they observe, act, and communicate with the environment and other agents. In cases where an agent's initiative is based on non-random internal stimuli, we categorize this as being \underlineColor{colorConfigurationLogic}{Internal Deliberate}, as the agent can decide when to take observations, actions, and communicate based on its own logic and world model, without external communication or instructions. In case the agent can take initiative by itself, but it does so based on a random variable, then we categorize this as being \underlineColor{colorConfigurationLogic}{Internal Spontaneous}. In contrast, when external actors take the initiative, we categorize this as being \underlineColor{colorConfigurationLogic}{External} autonomy. This label describes agents that only take initiative based on external communication, instructions from other agents, or the infrastructure layer.
Human agents are classified with \underlineColor{colorConfigurationLogic}{Internal Deliberate} autonomy by default, e.g., \cite{2024_Dash_DBromley, 2021_MI3_YZhang, 2010_PredictiveAnalyticsUsingBlackboard_JYu, 2021_Snowy_ASrinivasan}. This classification aligns with the human-centric nature of most of the analyzed systems and captures the freedom human users have in their choice to engage in VA tasks.
Jia et al. make use of the \underlineColor{colorConfigurationLogic}{Internal Spontaneous} autonomy in an active-learning agent to identify uncertain regions in the learnable space, which enables the agent to query the human agent based on these findings \cite{2022_TowardsVisualExplainableActiveLerning_SJia}. In high-stakes domains like finance, software agents lack autonomy, and the initiative is entirely with the human agent \cite{2016_TimeFork_SKarthik}. Furthermore, while others refer to an ongoing shift that increases the degree of autonomy of the software agents \cite{Holter2024}, most agents within the mixed-initiative VA systems only become active once they are triggered externally, e.g., \cite{2015_SeeDB_MVartak,2018_Podium_EWall,2012_InteractiveAnalysisOfBigData_JHeer,2019_UserbasedVAWorkflowForEMA_DCashman}.

\dimension{Adaptation}{colorConfigurationLogic}{
In what way is the agent adapting within the system?}{Internal Adaptation, External Observations, External Communication, None}

While interacting with the environment and other agents, agents might learn to optimize their performance on specific tasks through \underlineColor{colorConfigurationLogic}{Internal Adaptation}, learn the preferences of different agents, e.g., through communication, i.e., \underlineColor{colorConfigurationLogic}{External Communication}, learn to mimic the behavior of other agents, adapt based on changes in the environment (e.g., \underlineColor{colorConfigurationLogic}{External Observations}), or adjust in a way that uses available tools and resources to exert actions within the environment.  While some agents use these strategies to adapt to the current situation or excel in their primary task, others will maintain similar behavior throughout the entire session. In this case, these agents are encoded with the label \underlineColor{colorConfigurationLogic}{None}. Based on internal adaptation, agents can optimize their actions to improve their task performance, e.g., \cite{2025_InsigHTable_GLi, 2024_SupportingGuidedEVAOnTimeSeriesData_YShi, 2025_Sprout_YLiu} by measuring their own performance or learning from the identified data insights \cite{2012_InteractiveAnalysisOfBigData_JHeer, 2025_InteractiveDesignOfExperiment_RSplechtna, 2020_explAIner_TSpinner}. Furthermore, agents can use External Observations, e.g., to adapt their behavior when searching for data or making recommendations \cite{2024_GuidedByAI_SHa,2019_Viana_FSperrle}. External communication leverages agents' ability to learn the preferences of other agents, like in \cite{2021_LearningContextualizedUserPreferences_FSperrle}, where software agents learn the human agent's preferences.

\subsection{World Model \includegraphics[height=1em]{figures/AgentIcon.pdf}}

    With the World Model dimension, the design space aims to capture an agent's internal interpretation and representation of the holistic system, encompassing the environment and all other agents. Within the World Model, we describe the aspects of the system that an agent internalizes, the persistence of this internalization, and the agent's awareness within the system. Being aware of the data that the world model tracks and how long it persists is crucial to examining the abstract representation of the environment and for knowing what information is considered by the agent when making decisions. Furthermore, categorizing data, agent, and task awareness provides a valuable approach to reasoning about the agent's understanding of the environment, other actors, and the current tasks in the system. 

\dimension{Provenance Tracking}{colorWorldModel}{
What types of provenance are part of the agent’s world model?}{Observation History, Action History, Interaction History, Communication History, None}
  
This category describes the types for which provenance -- the record of past activities and exchanges -- is captured by an agent. \underlineColor{colorWorldModel}{Observation History} refers to the agent’s ability to keep a log of observations made within the environment. This is beneficial when agents need to track the environment for specific events or changes, and summarize these \cite{2012_VAAgentBasedFrameworkForIntrutionAlertAnalysis_RShittu}. Through the \underlineColor{colorWorldModel}{Action History}, an agent is able to capture its explicit manipulations or analytic operations exerted onto the environment. Current approaches utilize this to build on previous actions \cite{2020_DivingInsights_AMcNutt, 2023_DiverseInterationRecommendationForPublicUsers_YLi} or to make use of this knowledge during adaptation \cite{2025_InsigHTable_GLi, 2021_LearningContextualizedUserPreferences_FSperrle, 2025_Sprout_YLiu}. The action history keeps track of actions performed by an agent itself, whereas the \underlineColor{colorWorldModel}{Interaction History} captures the sequence of actions taken by other agents in the system. This enables agents to, e.g., track human agent interactions during data wrangling, and summarize these for further automation \cite{2025_Dango_WHChen}. \underlineColor{colorWorldModel}{Communication History} describes agents that keep track of any incoming or outgoing signals, enabling agents to take these into account when responding \cite{2025_Leva_YZhao} or work on a follow up task \cite{2024_InkSight_YLin,2016_Eviza_VSetlur}. In contrast, some agents employ no provenance tracking, where past activity is neither recorded nor made accessible to the agent (\underlineColor{colorWorldModel}{None}). 

\dimension{Persistence}{colorWorldModel}{To what degree is the World Model persistent?
}{Continuous, Session-Based, Not Persistent}

Persistence describes the longevity of the provenance tracking contained within the World Model of an agent. Some agents keep and extend their world model \underlineColor{colorWorldModel}{Continuous}ly, allowing them to retain any knowledge gained across multiple sessions. This allows agents, e.g., to continuously build on the current world model, enabling active learning scenarios \cite{2025_KMTLabeler_HWang, 2025_HEPHA_SZhou, 2018_VisIRR_JChoo}. Other agents only keep track of the provenance in a \underlineColor{colorWorldModel}{Session-Based} manner, meaning that their world model will never automatically consider states created within earlier sessions. E.g., agents that observe human interactions and build scripts based on this \cite{2012_InteractiveAnalysisOfBigData_JHeer, 2025_Dango_WHChen} or when monitoring another agent \cite{2019_Sherpa_ZCui} only require session-based persistence. While human agents have at least a session-based persistence, in some cases, software agents only retain the world model for one iteration or calculation between their observation and action. In such cases the world model is considered \underlineColor{colorWorldModel}{Not Persistent}. This can be the case for a summarizer agent that only needs the current observation to perform the summary action of a text document \cite{2023_ConceptEVA_XZhang}.

\dimension{Data Awareness}{colorWorldModel}{What is the degree of the agent's awareness of data present within the environment?
}{Full Data Awareness, Aligns with Observations, Subpart of Observation}

Within the world model, there exists a certain degree of data awareness of the total set of data within the environment. This category quantifies the amount of data the agent has access to within their world model. An agent may have \underlineColor{colorWorldModel}{Full Data Awareness}, in which case the agent has all data in the environment represented within their world model immediately upon starting a system, e.g., 
\cite{2022_SupportingSerendipitousDiscovery_MJasim,2025_WhatIF_AMishra,2023_Causeworks_TKapler}. We assign \underlineColor{colorWorldModel}{Aligns with Observations} to those agents that do not have full data awareness and build their data understanding based on their observations, like, e.g., \cite{2019_VAForTopicModeling_MElAssady,2024_Qutaber_QJiang,2025_Dango_WHChen}. In the event that an agent summarizes their observations and only builds a world model on this subset of the observations, we assign the label \underlineColor{colorWorldModel}{Subpart of Observation}, e.g., \cite{2018_Podium_EWall,2025_MixingLinters_AMcNutt,2011_GuidingFeatureSubsetSelection}.

\dimension{Agent Awareness}{colorWorldModel}{Is the agent aware of other agents within the system?
}{Aware of all Agents, Aware of a Subset of Agents, Not Aware of Others}

In mixed-initiative VA systems, by definition there must be at least two agents present. In certain systems, the agents are able to interact or communicate with one-another, while in other systems agents are exclusively interacting with the environment and not directly interacting with other agents. \underlineColor{colorWorldModel}{Aware of all Agents} is assigned when an agent is aware of all (other) agents within the system. This form of agent awareness allows an agent to position itself alongside other agents and adapt based on this, e.g., \cite{2023_Datatales_NSultanum,2024_DiscipLink_CZheng}. \underlineColor{colorWorldModel}{Aware of a Subset of Agents} describes agents that are aware of at least one other agent within the system. In many scenarios this applies to agents that execute tasks when instructed by another agent, e.g., \cite{2025_LightVA_YZhao,2025_Dango_WHChen}. In certain cases, agents are \underlineColor{colorWorldModel}{Not Aware of Others}, and do not position themselves relative to other agents within the system, like in \cite{2025_HEPHA_SZhou,2012_VAAgentBasedFrameworkForIntrutionAlertAnalysis_RShittu, 2021_MI3_YZhang}.

\dimension{Task Awareness}{colorWorldModel}{
Is an agent aware of its own tasks, and is the agent aware of the tasks of other agents?}{Self Aware, Aware of other Agents Tasks, No Awareness}

Task awareness factors in an agent's knowledge and reasoning capability regarding its own tasks and those of other agents. If an agent is aware of its own tasks, then we refer to this as the agent being \underlineColor{colorWorldModel}{Self Aware}; in this case an agent is able to reason and act with a sense of purpose. With this knowledge the agent can also adapt its behavior in case the task changes, e.g., \cite{2023_Datatales_NSultanum,2024_Dash_DBromley,2025_Leva_YZhao}. In the event an agent is aware of at least one other agent's tasks, then it is considered \underlineColor{colorWorldModel}{Aware of Other Agents' Tasks}. In this case an agent is able to reason and act with knowledge of the purpose of other agents, e.g., allowing the agent to provide better recommendations to these other agent \cite{2023_DiverseInterationRecommendationForPublicUsers_YLi,2019_DoWhatIMean_MTroy}. If neither of these codes apply, then an agent is considered to have \underlineColor{colorWorldModel}{No Awareness}. This does not mean the agent does not know how to execute its tasks, but refers to not being able to reason about them. E.g., creating rankings based on an initial list is a fixed algorithm that does not require any task awareness \cite{2018_Podium_EWall}.

\subsection{Observations \includegraphics[height=1em]{figures/AgentIcon.pdf}}
In the Observations dimension, we detail an agent's external perception -- all perceived information -- of the environment, categorizing what it can observe, when it chooses to observe, and how it interacts observationally with other agents. Understanding this dimension is crucial, as it shapes an agent's decision-making, responsiveness, and awareness, directly influencing the actions it can undertake within its environment. Furthermore, this dimension enables the comparison of different agents based on the amount and detail of information they can perceive.

\dimension{Data Structure}{colorObservation}{When observing the environment, how is the observed data structured?
}{Structured, Semi-Structured, Unstructured}

This category determines the structure of the observed data. When observing data that is presented and stored with a predefined structure, such as tables or visualizations, we consider this \underlineColor{colorObservation}{Structured} data (e.g., tables \cite{2007_Tibor_DLiu,2019_Fairvis_AACabrera}). In contrast, data with no predefined structure is considered \underlineColor{colorObservation}{Unstructured} data, e.g., plain text \cite{2019_Viana_FSperrle,2024_ConceptInduction_MSLam}. In case the data is, e.g., available in a JSON format, this is encoded as \underlineColor{colorObservation}{Semi-Structured}, e.g., \cite{2024_Dash_DBromley, 2025_HEPHA_SZhou}. Agents can be capable of observing multiple inputs, therefore multiple labels may apply to describe all input formats that an agent takes.

\dimension{Sensor Modality}{colorObservation}{
What sensory modalities are part of the agent's observation?}{Text, Numeric, Visual, Code}

In addition to the structure of the data, it is also valuable to describe what modalities of observed data the agent takes. While the structure of the data describes the format that the data takes, the sensor modality addresses the type of input data that the agent perceives. We refer to \underlineColor{colorObservation}{Text} for any textual data that the agent perceives, such as strings or characters, e.g., \cite{2021_ArchiText_HKim,2023_Datatales_NSultanum}. \underlineColor{colorObservation}{Numeric} describes integers, floats, vectors, or matrices, e.g., \cite{2016_TimeFork_SKarthik,2025_InteractiveDesignOfExperiment_RSplechtna, 2015_MixedInitiativeVAUsingTaskDrivenRecommendations_KCook}. Given the VA scope, \underlineColor{colorObservation}{Visual} describes any form of visual elements that are present within the UI of the VA system, e.g., images (\cite{2022_VideoModerator_TTang}), visualizations, charts (\cite{2016_Voyager_KWongsuphasawat}), etc. Finally, \underlineColor{colorObservation}{Code} is a textual modality that holds a set of instructions that can be executed by an agent or an interpreter within the environment, e.g., \cite{2021_QuestionComb_RSevastjanova,2025_Dango_WHChen}.

\dimension{Context Level}{colorObservation}{What part of the environment is observed by the agent?
}{One Input, Multiple Inputs, Whole Environment}

In most cases agents can only observe a fraction of the whole environment. This category -- Context Level -- addresses this by grouping the scope of the observations into three comprehensive classes. \underlineColor{colorObservation}{One Input} refers to all cases in which an agent only observes one singular section of the environment. This would be a single value, such as a vector indicating which elements have been selected by another agent, e.g., \cite{2025_GuidanceSourceMatters_ANarechania}. Usually this results in an agent having a small information window relative to the total set of available data within the environment. In cases where an agent observes multiple sections of the environment -- but not the full environment -- we describe this as the agent observing \underlineColor{colorObservation}{Multiple Inputs}, giving the agent more detailed insights into the available data, e.g., \cite{2010_Articulate_YSun,2021_LearningContextualizedUserPreferences_FSperrle}. An agent might be able to observe the \underlineColor{colorObservation}{Whole Environment}, allowing the agent to view all data in the environment. Thus, the agent has access to the full scope of the environment, which it can potentially integrate into its world model, e.g., \cite{2012_InteractiveAnalysisOfBigData_JHeer,2023_Causeworks_TKapler, 2016_Voyager_KWongsuphasawat}.

\dimension{Synchronism}{colorObservation}{
Is the environment able to change between an agent observing and consequently acting based on said observation?}{Synchronized, Asynchronized}

This category captures whether the environment can update between the moment of the observation and the moment the related action takes place. This is relevant for agents that base their actions on the most recent state of the environment to ensure better conformance. Agents that observe a state of the environment, and consequently execute their action before the state of the environment can change, are considered \underlineColor{colorObservation}{Synchronized}. These agents will always perform actions considering the latest state of the environment, like e.g, \cite{2024_iToT_ABoyle, 2016_Eviza_VSetlur}. In case the environment can change between an observation and action, we describe the agent's observations as being \underlineColor{colorObservation}{Asynchronous}, e.g., \cite{2025_Leva_YZhao, 2019_Viana_FSperrle,2021_InterwarvingMultimodalInteraction_ASrinivasan}.
In more detail this is the case if there are a software and a human agent and both are able to take actions in parallel while one agent has not finished its execution \cite{2025_LightVA_YZhao}.

\dimension{Trigger}{colorObservation}{
What triggers the observations of an agent?}{Internal, Communication Based, Infrastructure Based}

The trigger refers to the causal event that initiates an agent's observations. While this is related to an agent's autonomy within the Configuration and Logic dimension, it merits its own category, as it describes the event trigger that causes an agent to make an observation. To properly take part as an agent within a mixed-initiative VA system, the agent is required to make observations in order to contribute to the shared analytics task. If an agent can decide to observe the environment through an internal stimulus, like an internal calculation, we label this as an \underlineColor{colorObservation}{Internal} trigger, e.g., \cite{2021_MI3_YZhang, 2019_DoWhatIMean_MTroy}. An observation can also be caused by instructions from another agent, in which case the agent uses a \underlineColor{colorObservation}{Communication Based} trigger, e.g., \cite{2018_Podium_EWall, 2015_SeeDB_MVartak}. The final label \underlineColor{colorObservation}{Infrastructure Based} trigger describes when the infrastructure layer is responsible for causing an agent to observe, e.g., \cite{2025_Divisi_VSivaraman,2019_Fairvis_AACabrera, 2022_ChartSeer_JZhao}.

\subsection{Communication \includegraphics[height=1em]{figures/AgentIcon.pdf}}

The Communication dimension encompasses agent-to-agent interactions and signals that do not pass through the environment. It defines an agent's role in communication as well as the content of the messages that agents send and receive. By characterizing communication in greater detail—such as the payload, types of incoming and outgoing messages, and triggers—it becomes possible to analyze and compare the quantity and nature of information an agent receives from other agents. Additionally, this dimension outlines the internal sharing category, which allows for an examination of how much internal knowledge is shared among agents. 

\dimension{Payload}{colorCommunication}{
Is there any form of communication from an agent, and does the signal include a payload?}{Signal, Signal and Data, None}

The payload describes an important component of the agent-to-agent communication as it describes if there is any form of communication from an agent, and whether that communication includes data, i.e., a payload. We consider \underlineColor{colorCommunication}{Signal}, when an agent is capable of communication, but does not transmit any data in the process. E.g., agents can use signals to notify other agents about having finished their task or instruct other agents \cite{2021_QuestionComb_RSevastjanova, 2018_Podium_EWall, 2019_VAForTopicModeling_MElAssady}.  \underlineColor{colorCommunication}{Signal and Data} is instead assigned to describe agents who are capable of both communication and data sharing associated with the signals. The signals themselves only relate to the semantics of the communication, while the data refers to the information presented alongside a signal, e.g., a request (signal) is sent to transform a specific part of data provided by the requesting agent \cite{2024_Talk2Data_YGuo}. For those agents that are not capable of communicating, we classify them with the label  \underlineColor{colorCommunication}{None}, like, e.g., \cite{2024_InterpretingHighDimensionalProjectionsWithCapacity_YZhang, 2011_GuidingFeatureSubsetSelection, 2019_EvaluatingPreferenceCollection_CKuhlman}.

\dimension{Type Outgoing}{colorCommunication}{What are the signal types of the outgoing communication from an agent?
}{Notification, Instruction, Request, Feedback, Availability, None}

This category extends the payload category by describing the type of communication that an agent is sending. It clarifies the intent and purpose of an agent's communication.
\underlineColor{colorCommunication}{Notification} is assigned when an agent informs another agent, e.g., \cite{2022_IntentVizor_GWu, 2020_SemanticConceptSpaces_MElAssady}. \underlineColor{colorCommunication}{Instruction} describes an agent prompting another agent to take a certain action in a one-directional sense, i.e., the other agent does not communicate anything in return, like in \cite{2025_Sprout_YLiu, 2024_TopicRefiner_HYang}. \underlineColor{colorCommunication}{Request} refers to a bi-directional mode of communication where an agent asks another agent to take a certain action and to be informed of the result, e.g., the other agent has to search for specific data points \cite{2016_Eviza_VSetlur}. \underlineColor{colorCommunication}{Feedback} describes a specific form of notification, where an agent does not have the immediate intent of informing but rather to provide commentary related to another agent's actions, i.e., feedback \cite{2021_LearningContextualizedUserPreferences_FSperrle}, recommendations \cite{2019_Viana_FSperrle}, etc. Lastly,  \underlineColor{colorCommunication}{Availability} is assigned if an agent indicates its operational availability, i.e., busy or open, e.g., \cite{2019_DoWhatIMean_MTroy}. 

\dimension{Type Incoming}{colorCommunication}{
What are the signal types of the incoming communication to an agent?}{ Notification, Instruction, Request, Feedback, Availability, None }

Complementary to the Type Outgoing, we define the Type Incoming, enabling a comprehensive overview of the communication channels throughout the system. Logically, while one agent sends a specific type, at least one other agent within the same system receives this signal type. Note that the communication types of Type Incoming are the same as those of Type Outgoing.

\dimension{Trigger}{colorCommunication}{What trigger type is initiating the agent-to-agent communication?
}{Internal, Context Based, Communication Based, Infrastructure Based, None}

Similarly to the observation trigger, the cause for an agent sending a signal to another agent is considered the communication trigger. \underlineColor{colorCommunication}{Internal} describes an agent utilizing an internal stimuli to decide when to communicate with another agent, e.g., \cite{2024_Socrates_GWu,2021_LearningContextualizedUserPreferences_FSperrle}. \underlineColor{colorCommunication}{Context Based} refers to an observational trigger that causes communication to happen, such as information changing within the UI  \cite{2025_WizzardGuidanceStrategiesAndDynamics_FSperrle}. \underlineColor{colorCommunication}{Communication Based} describes direct triggers that arise from agent-to-agent communication. We usually observe this when requests or instructions are sent, like in \cite{2025_LightVA_YZhao}. \underlineColor{colorCommunication}{Infrastructure Based} categorizes triggers that are initiated by the Infrastructure layer, e.g., when communication happens right after an action is performed by an agent \cite{2016_Voyager_KWongsuphasawat, 2021_QuestionComb_RSevastjanova}. Agents that do not communicate, are labeled with \underlineColor{colorCommunication}{None}.

\dimension{Internal Sharing}{colorCommunication}{Is an agent sharing its world model and logic with another agent?
}{Operational Sharing, World Model Sharing, None}

What agents share about their internal status can be grouped into two classes. \underlineColor{colorCommunication}{Operational Sharing} refers to all information an agent provides to other agents regarding their internal logic. Examples of this include sharing an agent's internal weights or precise decision-making steps, e.g., \cite{2020_explAIner_TSpinner, 2021_IntegratingPriorKnowledge_APister, VideoPro_He_2024}. The other class is \underlineColor{colorCommunication}{World Model Sharing}, which generally revolves around an agent sharing their internalized interpretation and representation of the environment, e.g., this could include human agents who describe their understanding of a piece of text or a software agent that provide their understanding of a human user's preferences \cite{2018_Podium_EWall}. \underlineColor{colorCommunication}{None} is then utilized for those agents that do not partake in operational or world model sharing.

\subsection{Actions \textbf{\includegraphics[height=1em]{figures/AgentIcon.pdf}}}

The Actions dimension aims to illustrate how an agent interacts with its environment through external acts. This encompasses the various types of actions an agent can take, the modality of those actions, and the timing of their execution. Additionally, we consider the extent to which the environment is influenced by the agent's actions. This enables us to analyze the complexity of the agent's operational capabilities, the impact of the agent on environmental changes, and the timing of the actions the agent decides to carry out.

\dimension{CRUD Type}{colorActions}{What is the type of action taken by an agent?
}{Create, Update, Delete}

This category captures all high-level formats of actions that an agent can exert in the environment. CRUD refers to create, read, update, and delete. Generally, any action from an agent will either \underlineColor{colorActions}{Create} and add new data into the environment, e.g., \cite{2019_Viana_FSperrle, 2023_Datatales_NSultanum, 2024_CalliopeNet_QChen}, \underlineColor{colorActions}{Update} and change existing data within the environment, e.g., \cite{2025_WhatIF_AMishra, 2024_TopicRefiner_HYang, 2024_Stile_SKabir} or \underlineColor{colorActions}{Delete} and remove existing data from the environment, e.g., \cite{2021_ArchiText_HKim,2022_MutliVision_AWu,2012_VAAgentBasedFrameworkForIntrutionAlertAnalysis_RShittu}. While the term CRUD also includes read, this is already described within the observations dimension, and thus not a code within actions. Finally, by definition, each agent will have at least one CRUD type, as an agent is assumed to take action within the environment to be considered an agent. 

\dimension{Type}{colorActions}{
What is the modality of the action taken by an agent within the environment?}{Writing, Counting, Coding, Visualizing, Interacting}

Within the design space, we distinguish between the modalities of the data being manipulated by an agent within the environment, similarly to the observations. \underlineColor{colorActions}{Writing} refers to an agent creating, updating, or deleting textual information within the environment, such as strings or characters, e.g., \cite{2023_Datatales_NSultanum, 2024_CalliopeNet_QChen, 2025_WhatIF_AMishra}. In contrast, \underlineColor{colorActions}{Counting} describes general manipulations regarding integers, floats, vectors, and matrices, e.g., \cite{2022_HowDoAlgorithmicFairness_CRares} and, e.g., in case of an agent labeling medical data \cite{2025_KMTLabeler_HWang}. When manipulated data is used as executable instructions, then this is described with the label \underlineColor{colorActions}{Coding}, as in \cite{2025_Sprout_YLiu, 2025_Dango_WHChen,2025_LightVA_YZhao}. If the action produces a graph, chart, or other visualization within the UI, then this is described as \underlineColor{colorActions}{Visualizing}, e.g., \cite{2017_PrivacyPreserving_JKChou,2006_InteractiveVisualSynthesisOfAnalyticalKnowledge_DGotz,2022_MutliVision_AWu}. Lastly, if agents are exclusively manipulating the UI, then this is considered \underlineColor{colorActions}{Interacting}, e.g., \cite{2024_TopicRefiner_HYang, 2024_Qutaber_QJiang}. Due to the nature of mixed-initiative systems and humans' roles in them, most human agents will have this action type.  

\dimension{Affected Scope}{colorActions}{
What part of the environment and world model is affected by an action?}{Entire Environment, Part of World Model, Outside of World Model}

When manipulating the environment, each action by an agent affects a different aspect of the environment. An action can affect the \underlineColor{colorActions}{Entire Environment}, and have an effect on all data within the system, such as deletion of an entire dataset, e.g., \cite{2007_Tibor_DLiu, 2025_InteractiveDesignOfExperiment_RSplechtna}. For those actions that do not affect the whole environment, a destinguishment is made between; \underlineColor{colorActions}{Part of World Model}, where an agent takes an action that affects a part of the environment that is within their world model, e.g., \cite{2012_VAAgentBasedFrameworkForIntrutionAlertAnalysis_RShittu,2025_Dango_WHChen, 2025_Sprout_YLiu}, and \underlineColor{colorActions}{Outside of World Model}, in which case an agent affects a part of the environment that is not within their world model, e.g. by creating a visualization or recommendation in an unselected area by the user \cite{2025_MixingLinters_AMcNutt, 2016_Voyager_KWongsuphasawat,2025_RSVPForVPSA_MKlaffenboeck}.

\dimension{Trigger}{colorActions}{
What triggers an agent's action to happen?}{Internal, Context Based, Communication Based}

Similar to the communication and observation trigger categories, agents also have a certain trigger that initiates their actions. \underlineColor{colorActions}{Internal} refers to actions that are triggered by stimuli within the agent, such as an agent that periodically removes all visualizations within the UI, e.g., \cite{2019_Viana_FSperrle, 2025_Dango_WHChen, 2024_iToT_ABoyle}. \underlineColor{colorActions}{Context Based} instead means that an action is a direct result from an observation. This is inherently the label for agents whenever they do not have a context-based trigger for communication, e.g., \cite{2015_MixedInitiativeVAUsingTaskDrivenRecommendations_KCook,2025_MixingLinters_AMcNutt,2007_Tibor_DLiu}. Actions can also be caused through the reception of \underlineColor{colorActions}{Communication Based} signals, such as instructions or requests, e.g., \cite{2025_Dango_WHChen, 2025_HEPHA_SZhou, 2025_Sprout_YLiu}. Agents, by definition, must take at least one action. Therefore, a definition of a "none" label would be redundant.

\dimension{Delay}{colorActions}{
When an agent takes an action, is the action delayed or executed immediately?}{No Delay, Delay with Static Duration, Delayed Based on a Metric, Delayed until Action Acknowledged by Other Agent}

In mixed-initiative, where multiple agents interact, it is crucial to determine the optimal timing of actions in a sequence of effects to effectively progress towards the task goal. Generally, when taking actions, agents can decide to delay a given action before putting the result in effect. This category describes these delays. \underlineColor{colorActions}{No Delay} refers to actions that are immediately put into the environment, e.g., an agent that is triggered to take an action and immediately executes it \cite{2007_Tibor_DLiu, 2018_Podium_EWall,2021_LayoutExOmize_PSchader}. \underlineColor{colorActions}{Delay with Static Duration} indicates actions that are taken within a fixed interval. While in other cases, agents may calculate the best delay to execute an action, in which case the action is \underlineColor{colorActions}{Delayed Based on a Metric}, e.g., \cite{2020_DivingInsights_AMcNutt}. Finally, agents may want to confirm with another agent before executing. In these cases, the action is \underlineColor{colorActions}{Delayed until Action Acknowledged by Other Agent}, e.g., requesting further clarification \cite{ 2025_Dango_WHChen} or proposing an action to a steering agent \cite{2021_LearningContextualizedUserPreferences_FSperrle}. 

\subsection{Infrastructure \includegraphics[height=1em]{figures/InfrastructureIcon.pdf}}

With the Infrastructure dimension, our design space seeks to explain the general steering unit of the system itself, which cannot be encapsulated within an agent. Furthermore, the Infrastructure dimension includes some metadata regarding the holistic system. We describe within the Infrastructure what modules are initialized, what modules change dynamically, the agent interplay within the system, and the analytic tasks of the system. It's important to note that, in contrast to the agents, the percentages in this section are derived from the total number of systems that encompass all agents.

\dimension{Initialized Modules}{colorInfrastructure}{What modules are initialized within the infrastructure?
}{Configuration Module, Observation Module, Action Module, Communication Module}

Within the infrastructure, multiple modules can be initialized based on the requirements of a system within the mixed-initiative VA context. This category specifically refers to any modules that are initialized when a system is started. The \underlineColor{colorInfrastructure}{Configuration Module} provides general setup parameters for all agents and the environment, e.g., \cite{2025_Dango_WHChen,2021_LearningContextualizedUserPreferences_FSperrle,2025_LightVA_YZhao}. While the \underlineColor{colorInfrastructure}{Observation Module} and \underlineColor{colorInfrastructure}{Action Module} initialize the observation (e.g., \cite{2022_HowDoAlgorithmicFairness_CRares, 2018_Podium_EWall, 2025_ScholarMate_RYe}) and action units within the agents respectively (e.g., \cite{2025_Sprout_YLiu,2024_InkSight_YLin,2021_GeonoCluster_SDas}). The \underlineColor{colorInfrastructure}{Communication Module} is setup to facilitate agent-to-agent communication to exchange information, e.g., \cite{2023_Datatales_NSultanum, 2024_iToT_ABoyle, 2025_Dango_WHChen}. Generally, within the mixed-initiative setup, the majority of modules are initialized. However, in rare cases, one or more modules are not necessary. 

\dimension{Dynamic Modules}{colorInfrastructure}{
What models are dynamically configured by the infrastructure layer during run-time?}{Configuration Module, Observation Module, Action Module, Communication Module, None}

Over the course of a session, the initialized modules may adapt dynamically within the infrastructure, allowing the initial initialization to evolve over time. The \underlineColor{colorInfrastructure}{Configuration Module} allows the infrastructure layer to change the behavior logic of each agent, and the environment. Changing the \underlineColor{colorInfrastructure}{Observation Module}, e.g., \cite{2024_DiscipLink_CZheng}, and \underlineColor{colorInfrastructure}{Interaction Action Module} can dynamically update the observation- and action behavior of agents, respectively. Lastly, updating the \underlineColor{colorInfrastructure}{Communication Module} can change the communication behavior and adjust which agents are communicating with each other. For those systems without any dynamic models, we coded these with \underlineColor{colorInfrastructure}{None}, e.g., \cite{2025_VisCars_PMoens,2020_DivingInsights_AMcNutt, 2007_Tibor_DLiu}.

\dimension{Trigger of Dynamic Modules}{colorInfrastructure}{What triggers the dynamic adjustment of the modules?
}{Planned Time Steps, Based on Agents, Based on Environment, None}

Based on the dynamic modules that are initialized, we also categorize how these dynamic modules update within the mixed-initiative VA system. The modules can be dynamically adjusted over predefined intervals using \underlineColor{colorInfrastructure}{Planned Time Steps}. In certain settings, the infrastructure may also dynamically update the modules \underlineColor{colorInfrastructure}{Based on Agents} within the system, e.g., \cite{2024_DiscipLink_CZheng}, or alternatively \underlineColor{colorInfrastructure}{Based on the Environment}, as specific states could trigger the dynamic changes. For those systems where no dynamic modules are initialized, we coded these systems with \underlineColor{colorInfrastructure}{None}, e.g., \cite{2025_VisCars_PMoens,2020_DivingInsights_AMcNutt, 2007_Tibor_DLiu}.

\dimension{Agent Interplay}{colorInfrastructure}{
What is the nature of the relationship of the interplay between agents within the mixed-initiative VA system?}{Independent, Cooperative, Competitive}
    
As the scope of the design space includes mixed-initiative systems, it is important to describe how agents interact in a shared system. Agents may interact in an \underlineColor{colorInfrastructure}{Independent} manner, where each agent works on completing their own task to contribute to an overall goal, e.g., \cite{2024_InkSight_YLin,2025_InsigHTable_GLi,2020_DivingInsights_AMcNutt}. Alternatively agents can engage in a \underlineColor{colorInfrastructure}{Cooperative} setting, where agents contribute to and solve tasks together inline with the overall goal of the system, e.g., \cite{2025_HEPHA_SZhou, 2021_QuestionComb_RSevastjanova,2020_HumanSteerableAI_AGonzalezMartinez, 2025_KMTLabeler_HWang}. In contrast, agents can interact in a \underlineColor{colorInfrastructure}{Competitive} setting, where, based on shared tasks, agents try to compete for the best performance, or try to learn another agent's preference, e.g., \cite{2021_LearningContextualizedUserPreferences_FSperrle}.

\section{Findings}
In this section, we elucidate the salient trends uncovered during our analysis, with a particular emphasis on the dynamics and interrelations among individual dimensions. First, we present findings and trends that we identified at the system level, followed by a more detailed analysis at the agent level.

\subsection{System Findings}

\begin{wrapfigure}[11]{r}{.55\textwidth}
    \vspace{-1.5em}
            \centering
    \includegraphics[width=.55\textwidth]{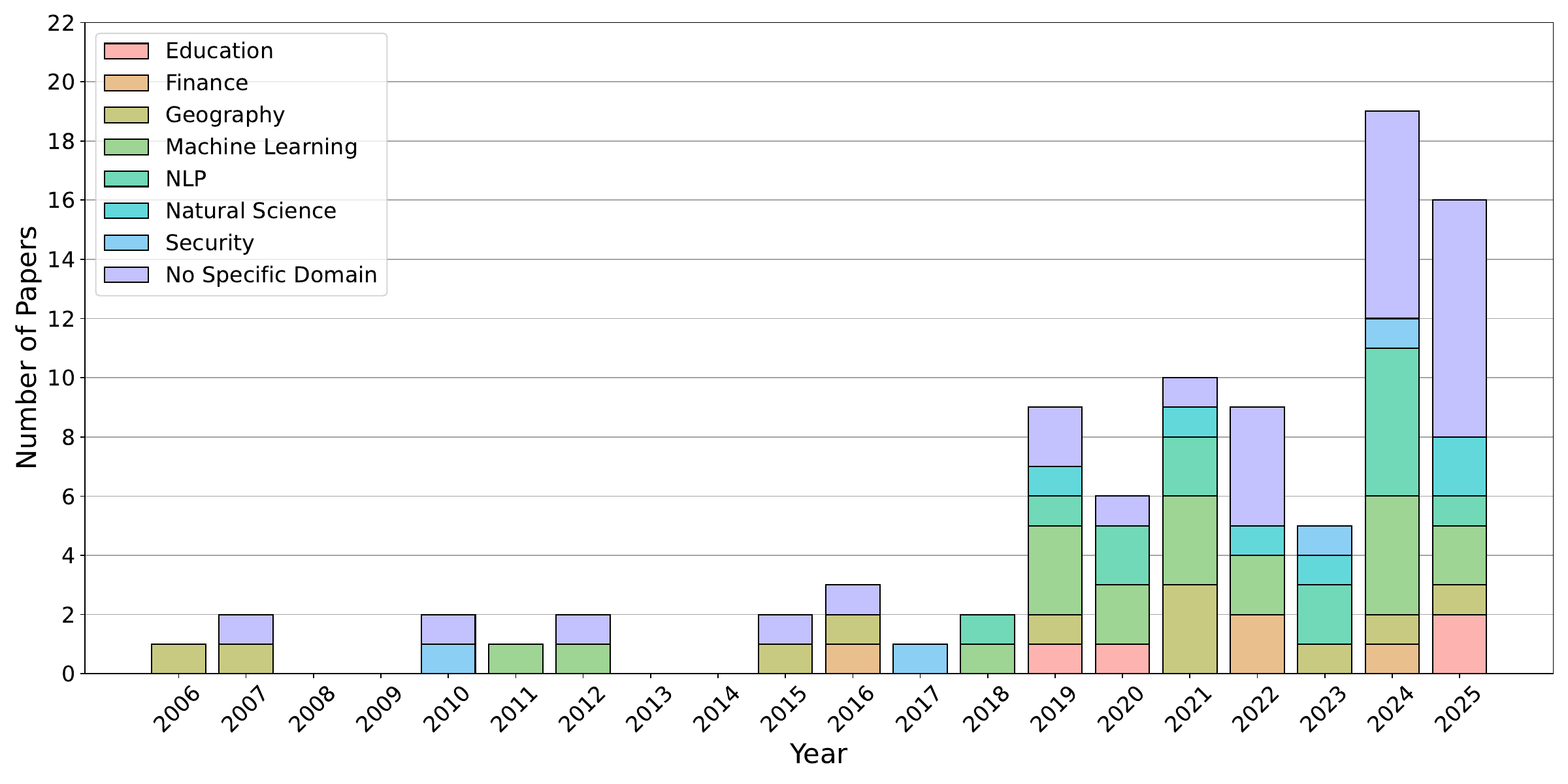}    \vspace{-3em}
    \caption{Papers in the scope of the design space by year and application domain. Our scope encompasses a wide range of systems across various fields, which have recently gained significant popularity.}
    \Description{The figure shows the number of papers per year that have been included within the design space. There is a gradual increase over the years, with particularly high increases in 2019 and 2024. The scale shows papers published between 2006 and 2025, with the highest number of papers within a single year being 19, which occurred in 2024.}
    \label{fig:frequency_paper_over_years}
    \vspace{-2em}
    \end{wrapfigure}

Within the scope of this work, we examined 90 different mixed-initiative VA systems published between 2006 and 2025. \autoref{fig:frequency_paper_over_years} illustrates the distribution of papers per domain across the publication years, revealing clear peaks
in 2019 and 2024. Overall, there is an upward trend in the number of systems being published in this domain. Most likely, this stems from an interplay of increased technology advancements -- in detail in the field of machine learning, increasing data complexity and amount of data, and the advantages of mixed-initiative for balancing the degree of human control and automatization in times of black-box models. We further see a rise in domain-specific applications, which promisingly indicates the practical adoption of such systems across domains due to an increased request for analysis and decision-support systems that combine expert knowledge and AI advancements. 
When analyzing the agent interplay within each system, we found that only one system \cite{2021_LearningContextualizedUserPreferences_FSperrle} employs a competitive agent interplay approach. Most likely, this is due to the nature of mixed-initiative being built to enable the cooperation of human and software agents. In contrast, other agents are part of a system to accomplish a specific, individual task (individualists). However, in \autoref{sec:opportunities}, we describe the benefits that can arise using a competitive approach. In total, we identified 207 individual agents within all review systems, resulting in a total average of 2.27 agents per system. While it is common to have multiple software agents within a system, only two reviewed papers feature multi-human agent systems \cite{2012_VAAgentBasedFrameworkForIntrutionAlertAnalysis_RShittu, 2022_HowDoAlgorithmicFairness_CRares}. In addition, there is no trend of increasing or decreasing average number of agents within a system.

\subsection{Agent Findings}
\begin{figure}[h]
    \centering
    \includegraphics[width=\textwidth]{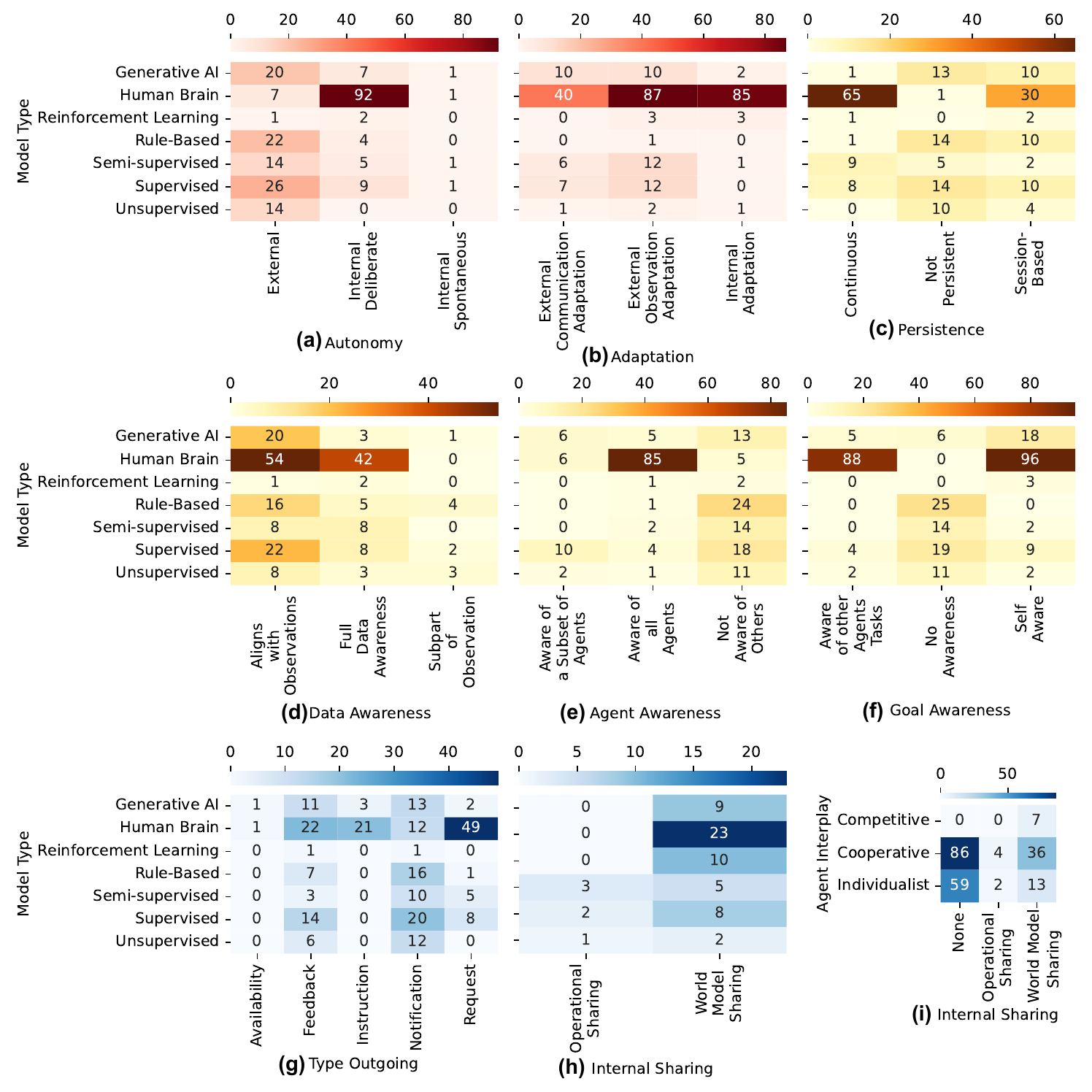}
    \caption{Correlations between the Model Types and other design space categories: (a) Autonomy, (b) Adaptation, (c) Persistence, (d) Data Awareness, (e) Agent Awareness, (f) Goal Awareness, (g) Communication Type Outgoing, (h) Status Sharing. (i) Shows the correlation of Agent Interplay and Internal Sharing.}
    \Description{The figure shows eight matrices, each indicating a different correlation between Model Type and another variable. Each matrix is described according to the corresponding variable: (a) Autonomy, which describes the autonomy, showing that generally the Human Brain has the highest autonomy, meaning that software agents generally have external autonomy. Additionally, internal deliberation does not often appear. (b) Adaptation describes the adaptation; it shows that the Human Brain tends to learn more often. Additionally, certain models tend to learn more often, namely, Generative AI, Semi-Supervised, Supervised, and Reinforcement Learning. Internal adaptation is rare among software agents. (c) Persistence describes persistence. Generally, the human brain has continuous persistence, which is rare for most other model types. Semi-supervised learning is the only exception, in that it is often continuous. (d) Data Awareness describes the data awareness. It shows that most model types have alignment with their observations, while in some cases, they are only aware of a sub-part of the observations, which is the case for rule-based and unsupervised models. The human brain then tends to have the highest full data awareness. (e) Agent Awareness describes the agent awareness; it shows that the human brain generally has high awareness, while this is rare for other model types, excluding generative AI.  Certain model types have better agent awareness, such as generative AI and Supervised, while other model types do not have a lot of agent awareness. (f) Goal Awareness: describes the goal awareness. The human brain is consistently self-aware, with many generative AI agents also being self-aware. The majority of model types have no goal awareness in most cases. Being aware of other agents' goals is rare, outside of the human brain. (g) Type Outgoing describes the type of outgoing communication, which shows a clear trend where the human brain makes requests, provides feedback, and issues instructions. The other model types tend to participate in notifying or providing feedback. This shows a similar trend to the autonomy, where the human brain more often has internal autonomy. (h) Status: describes the status. World Model Sharing is done most often by generative AI, reinforcement learning, supervised learning, and the human brain. Whilst Operational Sharing is rarely done, and mostly by the semi-supervised model type. (i) The matrix is displaying the number of agents within each combination of Internal Sharing and Agent Interplay. The majority of agents do not partake in internal sharing, while almost none take part in Operational Sharing. Finally, there is a clear increase in World Model sharing for cooperative systems.}
    \label{fig:findings_modeltype_cross_correlation}
\end{figure}

When analyzing the agent interplay within a system on an agent level, \autoref{fig:findings_modeltype_cross_correlation} (i), we noted that there is a larger number of agents sharing their world model than in the other interplay modes. This is most likely due to the idea of allowing the cooperation between the agents to be better aligned by sharing their world models. Note that within the competitive system, all agents are part of the same system; software agents are similarly modeled and compete for human attention, partly by sharing their world models.

\begin{wrapfigure}[7]{r}{.55\textwidth}
    \vspace{-2.5em}
   \centering
    \includegraphics[width=.55\textwidth]{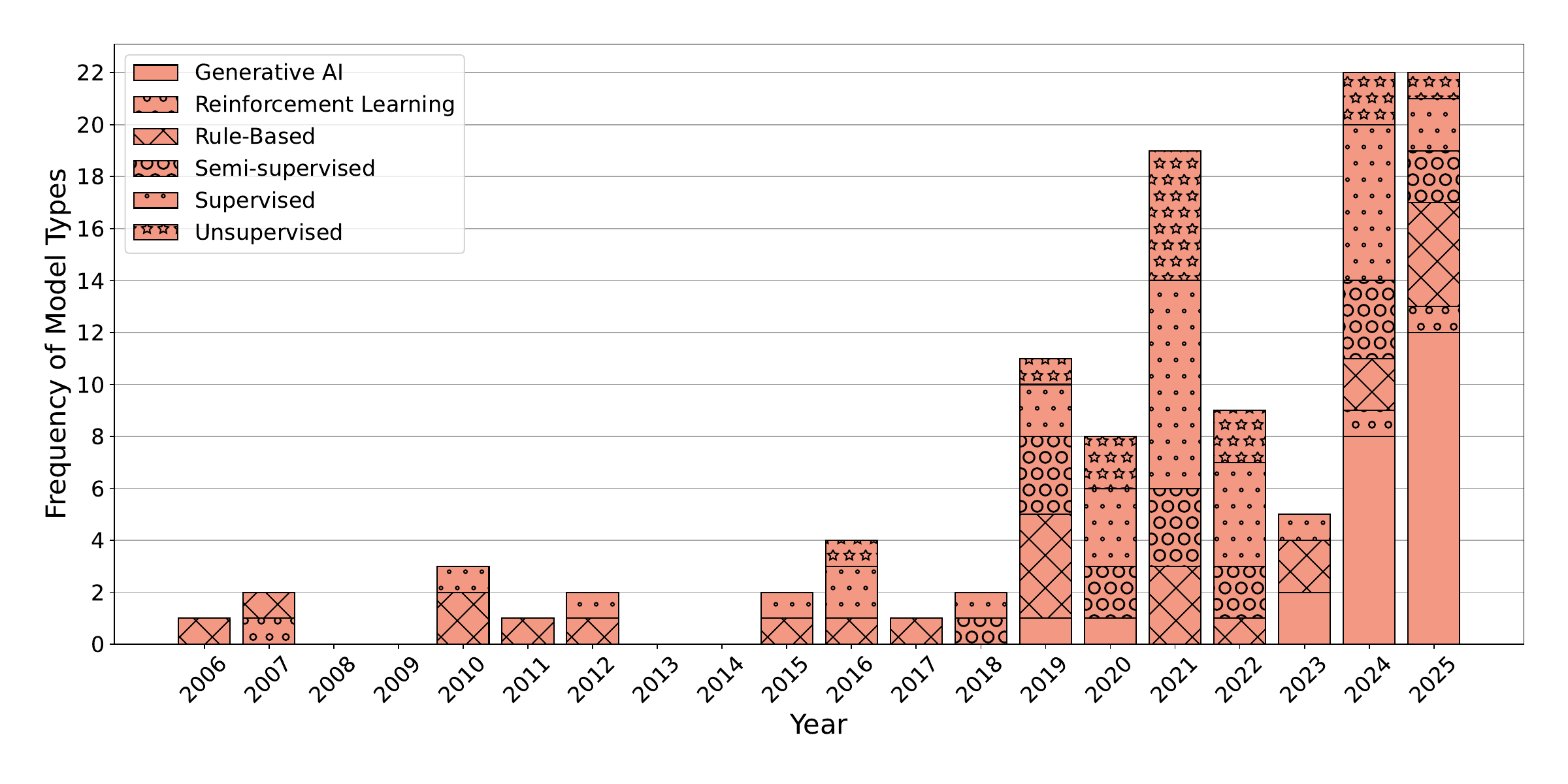}
    \vspace{-3.5em}
    \caption{Usage of Model Type by year}
    \Description{The figure shows the frequency of specific model types (excluding the Human Brain) per year. The model types shown are: Generative AI, Reinforcement Learning, Rule-Based, Semi-Supervised, Supervised, and Unsupervised. Rule-Based model type has been around the longest, and stays consistent in its usage over the years, while Generative AI was first used in 2019 and has become increasingly prominent over the years.}
    \label{fig:findings_modeltypes_per_year}
    \vspace{-2em}
    \end{wrapfigure} 
\paragraph{Comparing Agent Model Types} Advancements in machine learning within the past years have created strong contributions in different model types, especially in Generative AI. This trend can be quantitatively observed in \autoref{fig:findings_modeltypes_per_year}, which shows a surge in all machine learning types besides rule-based. Especially in the past two years, we have observed a boom in the use of Generative AI by agents.

\paragraph{Comparison on the Dimensions}
Analyzing the Autonomy and Adaptation categories in correlation with the model type (\autoref{fig:findings_modeltype_cross_correlation} (a + b)), we generally identified low autonomy in software agents, with no specific model type standing out in terms of autonomy. Moreover, internal adaptation is intrinsic to human agents but only appears sporadically in software agents (\autoref{fig:findings_modeltype_cross_correlation}, (b)).
Furthermore, within the World Model Dimension, semi-supervised models show 
\begin{wrapfigure}[11]{r}{.55\textwidth}
    \vspace{-2em}
      \centering
    \includegraphics[width=0.55\textwidth]{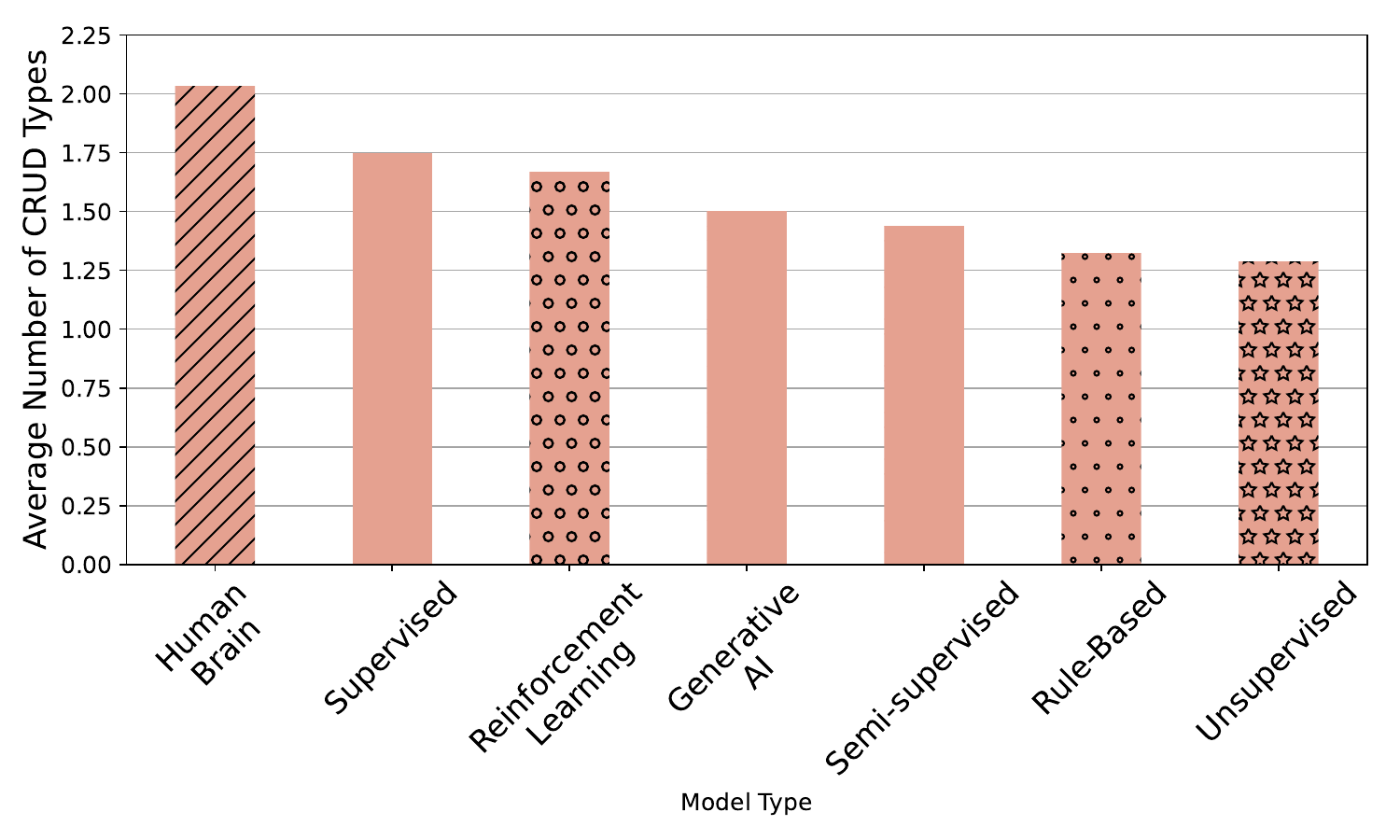}    
    \vspace{-3em}
    \caption{Average Count of the CRUD Set Size per Agent Model Type}
    \Description{The figure shows the average size of the set of CRUD Type values per model type. This is a measure of actionable diversity within the environment for the model type carried by an agent. Generally, the human brain has the largest average CRUD Type set (2.03), whilst the rule-based (1.32) and unsupervised (1.29) model types lag behind.}
    \label{fig:findings_CRUDpower}
    \vspace{-2em}
    \end{wrapfigure} 
strong persistence due to the focus on iterative training (\autoref{fig:findings_modeltype_cross_correlation}, (c)). Besides this, all software agent model types (Rule-based, Supervised, Semi-Supervised, Unsupervised, Reinforcement Learning, and Generative AI)  have weak awareness compared to their human brain counterparts when comparing data, agent, and task awareness (\autoref{fig:findings_modeltype_cross_correlation}, (d,e,f)). Within the communication dimension, currently requests and instructions are initiated mainly by human agents, while software agents rarely assume this role (\autoref{fig:findings_modeltype_cross_correlation}, (g)). In general, availability sharing is uncommon among all agents, with only two agents out of all reviewed agents \cite{2019_DoWhatIMean_MTroy,2025_LightVA_YZhao} (\autoref{fig:findings_modeltype_cross_correlation}, (g)). World model sharing is occasionally evident among humans, generative AI, and rule-based agents. In contrast, operational sharing is rare -- only six identified agents, none involving human or Generative AI (\autoref{fig:findings_modeltype_cross_correlation}, (h)). 
Regarding Actions, software agents tend to perform visualization actions more frequently compared to human agents. At the same time, there are only two cases of software agents that directly interact with UI elements within the user interface \cite{2024_TopicRefiner_HYang, 2024_Qutaber_QJiang}. Furthermore, comparing model types over the total set size of CRUD operations, the human brain has, on average, access to more CRUD operations (\autoref{fig:findings_CRUDpower}).

\paragraph{Findings in Infrastructure Dimension}
While almost all modules are always initialized at the start of each system session (in case of no communication, this module is not initialized), there is only one system that adapts a subpart of modules dynamically \cite{2024_DiscipLink_CZheng}. We did not observe any systems that make use of fully dynamical modules.

\paragraph{Summary} Comparing human agents and software agents in general, human agents demonstrate stronger awareness, persistence, adaptability, and autonomy than software agents, which at the moment makes human agents the dominant force in mixed-initiative VA systems, holding more power and responsibility. Software agents' contributions are more structured and less flexible, but they hold substantial potential for growth -- particularly through reinforcement learning, semi-supervised learning, and generative AI methods. Furthermore, while not yet being used and explored in much detail, as we show in the following opportunity section (\autoref{sec:opportunities}), dynamic adaptations of the infrastructure modules can provide benefits for the entire system.

\section{Opportunities}
\label{sec:opportunities}

Building on the previous findings, this section explores underrepresented dimensions and configurations of systems that suggest avenues for future research and design. 
First, we investigate how system-level augmentations and agent interactions can lead to benefits. Additionally, we examine how enhanced communication and awareness contribute positively, and how adaptive software agents combined with human-centric interaction patterns can facilitate improved mixed-initiative collaboration. 
Finally, we highlight opportunities arising from recent advances in Generative AI.
To illustrate these opportunities, we introduce a fictional example system, TreeHouse, which we reference throughout.

\begin{figure}[H]
    \centering
    \includegraphics[width=.9\textwidth]{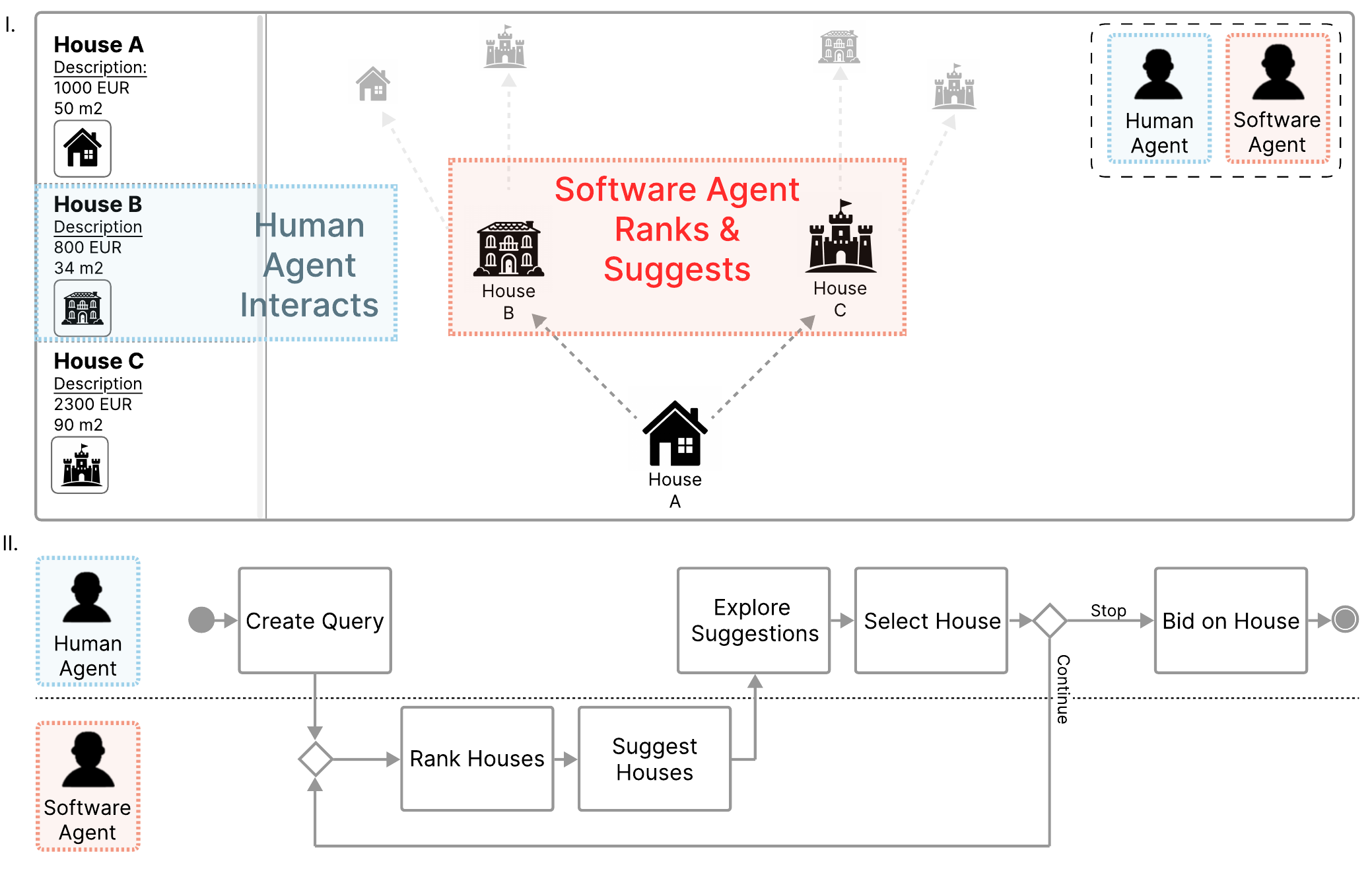}
    \caption{TreeHouse example, detailing a mixed-initiative visual analytics system utilizing exploration through a graph of recommended nodes. Shown through a I.) mock-up user interface and II.) workflow.}
    \label{fig:treehouse}
    \Description{The figure describes an example system within the mixed-initiative visual analytics domain. On top of a user interface is described, with a main view on the right that displays multiple houses that the software agent suggests in the form of a tree (graph) that fans out. Additionally, on the top right of this view, the agents within the system are indicated: Human and Software. Finally, for the user interface, a bar on the left is present, displaying a selection of available houses and descriptions from the current set of recommendations, allowing a user to click on these to get further recommendations. Below the user interface, a Unified Modeling Language activity diagram is shown. The activity diagram details how a human agent interacts with the system and how a software agent responds accordingly. The diagram starts with "Create Query" by the human, and then the software agent starts with "Rank Houses" and afterwards "Suggest Houses". Subsequently, the human agent will "Explore Suggestions" and "Select House". A gateway opens, as the Human can either stop and "Bid on House" or continue, in which case the software agent starts over with "Rank Houses".}
\end{figure}
\cbstart{\noindent\underline{\textbf{\textsc{TreeHouse} Example:}}  
TreeHouse is a fictional system in the mixed-initiative VA domain. It demonstrates both a potential system and a means of contextualizing the design space. The system assists users in navigating the housing market and identifying suitable homes by supporting tree-based exploration. It visualizes the user’s search path through a dataset of available properties, characterized by attributes such as size, cost, and location. A software agent monitors this exploration and provides suggestions aligned with the user’s predefined preferences. \autoref{fig:treehouse} shows a mock-up of the user interface and the workflow of the system.}

\cbend\

\subsection{Augmented System-Level Dynamics and Interaction}

In this subsection, we explore the augmentation of dynamics and interactions within mixed-initiative VA systems, highlighting the potential of competitive agent interplay, multi-user involvement, and dynamic system adaptation.

\begin{itemize}
    \item[\textbf{\texttt{Opp-1}}] 
\textsc{\textbf{Competitive systems can foster diverse perspectives, iterative learning, and enhance performance through strategic interaction} | } Based on our findings from the characterization of agents in existing systems, there are predominantly cooperative and individualist systems within the mixed-initiative VA domain. We identified only one system \cite{2021_LearningContextualizedUserPreferences_FSperrle} that utilizes competitive agent interplay. 
Competitive systems, however, could provide value by fostering diverse analytical perspectives across agents. Competition could also enable iterative refinement, as agents adapt to one another’s behavior, and improve overall system performance by driving higher quality and speed in analytic tasks.
\end{itemize}

\cbstart{\noindent\underline{\textbf{\textsc{TreeHouse} Example:}}
Applying these insights to an extension of TreeHouse, multiple software agents competitively generate recommendations and learn from user interactions. Their differing model types and optimization strategies provide varied suggestions, enabling the user to receive more refined recommendations.}\cbend\

\vspace{.7em}
\begin{itemize}
\item[\textbf{\texttt{Opp-2}}] 
\textsc{\textbf{Multi-user and multi-stakeholder systems leverage task distributions and different perspectives while covering expanded areas of analysis} | } Based on our findings, current systems mostly involve only one human agent per system.
Multi-user and multi-stakeholder systems, however, can leverage diverse perspectives and expertise while expanding the scope of possible tasks by distributing tasks among users. Collaborative efforts toward shared analytical goals could also enhance overall performance, enabling more efficient and comprehensive outcomes.
\end{itemize}
\vspace{.4em}
\cbstart{\noindent\underline{\textbf{\textsc{TreeHouse} Example:}} 
In another extension of TreeHouse, a second human user collaborates with the original user to explore housing options simultaneously. An additional software agent summarizes their shared progress, supporting more effective decision-making.}\cbend\

\vspace{.7em}
\begin{itemize}
\item[\textbf{\texttt{Opp-3}}] 
\textsc{\textbf{Dynamic systems can demonstrate adaptive capabilities through evolving agent roles and communication modules} | } We have identified only one system that has some form of dynamically updating modules \cite{2024_DiscipLink_CZheng}. This system only has a dynamic interaction module for observations, and all other dynamic modules are not present within the design space.
Dynamic modules can provide flexibility by supporting multi-stage systems where analytic task types and system behavior change over time in response to internal variables. Such models can also improve alignment with user preferences across evolving tasks and allow for the introduction of new agents or adjustments to existing roles. Moreover, they can strengthen information exchange by tailoring communication to specific tasks, specifications, and expertise, while enabling the redistribution of responsibilities between agents as needed.
\end{itemize}
\vspace{.4em}
\cbstart{\noindent\underline{\textbf{\textsc{TreeHouse} Example:}} 
Finally, in a last extension, the system monitors exploration progress and, once a threshold is reached, shifts to a ranking task by prompting the user to reflect on identified trends. At this stage, a new software agent appears to generate counterfactuals, ensuring deeper analysis. This dynamic adaptation creates a multi-stage system that adjusts to user progress and task demands.}\cbend\

\subsection{Enhanced Communication and Awareness}
Besides augmenting system dynamics and agent interplay, we further explore the pivotal role of enhanced communication and awareness within agent systems. Within our examples, we show how these reduce cognitive load, enhance user experience, and improve the coordination of the agents.

\vspace{.7em}
\begin{itemize}
\item[\textbf{\texttt{Opp-4}}] 
\textsc{\textbf{Availability sharing optimizes agent coordination, reducing cognitive load and enhancing focus} | } We observed only two agents that communicate their availability with other agents in the system \cite{2019_DoWhatIMean_MTroy,2025_LightVA_YZhao}. Both of these are software agents (one is a wizard), and there exists no system within mixed-initiative VA where a human user shares their availability. 
Availability sharing, however, could reduce system performance demands while strengthening agent-to-agent communication through readiness signals. Furthermore, users could gain greater control over agent behavior and experience less mental load, as the indication of availability would help to prevent overwhelming or distracting communication.
\end{itemize}    
\vspace{.4em}
\cbstart{\noindent\underline{\textbf{\textsc{TreeHouse} Example:}} In an extension of TreeHouse, the agent generates real-time suggestions based on user interactions, such as clicks and searches. To prevent excessive cognitive load, the interface incorporates a toggle that allows the user to control when feedback is displayed, maintaining focus on their primary task.}\cbend\

\vspace{.7em}
\begin{itemize}
\item[\textbf{\texttt{Opp-5}}] 
\textsc{\textbf{Status communication can enhance agent understanding and control through shared models and operational insights} | }  We notice that very few agents participate in status sharing, which includes both world model and operational sharing. Notably, we only identified six agents throughout the observed systems that partake in operational sharing. Status communication is important as it can enhance systems by enabling both world model and operational sharing. The sharing of world models can improve user understanding of agent knowledge and interpretation of the environment. In addition, it allows agents to learn from one another’s models and enables verification of shared knowledge. Operational sharing can also clarify agent behavior, highlight strengths and weaknesses, elucidate decision-making, and enhance human steering. For example, if a human agent understands that the software agent uses reinforcement learning and incorporates feedback from other agents to adapt its behavior, the human is more likely to provide high-quality feedback. With some additional knowledge about reinforcement learning, the human can also comprehend why the agent sometimes takes actions that are not the most probable ones.
\end{itemize}
\vspace{.4em}
\cbstart{\noindent\underline{\textbf{\textsc{TreeHouse} Example:}} 
TreeHouse example:
As another extension of the TreeHouse scenario, the human user rates agent recommendations on a scale of 1 to 5. The agent then reveals step-by-step reasoning that links these ratings to subsequent recommendation changes, clarifying how user feedback influences system behavior.}\cbend\

\vspace{.7em}
\begin{itemize}
\item[\textbf{\texttt{Opp-6}}] 
\textsc{\textbf{Enhanced agent- and goal awareness improves mixed-initiative systems in agent interplay, task execution, and explainability} | } Based on our observations, there exist a few software agents that are aware of other agents and their goals.
If an agent is aware of the presence and goals of other agents, it can adapt its behavior to avoid conflicting with those goals. For instance, in a scheduling system, if a software agent is aware that a human agent has the goal of keeping afternoons free for meetings, it can adapt its behavior by avoiding the automatic allocation of tasks during that time, thereby preventing a conflict of goals.
\end{itemize}
\vspace{.4em}
\cbstart{\noindent\underline{\textbf{\textsc{TreeHouse} Example:}} 
In a more advanced version of TreeHouse, multiple agents offer recommendations with complete knowledge of each other's goals and actions. This understanding facilitates seamless adaptation, task redundancy in case of failure, and enhanced overall system resilience.}\cbend\

\subsection{Adaptive Agents and Human-centric Interaction}

In exploring the evolving landscape of adaptive software agents, this collection investigates how dynamic adaptation enhances user integration, system usability, and agent initiative.

\vspace{.7em}
\begin{itemize}
\item[\textbf{\texttt{Opp-7}}] 
\textsc{\textbf{Adaptive software agents will demonstrate enhanced user integration, usability, and system initiative through dynamic adaptation} | } Based on our findings, we observe that software agents are rarely adaptive outside of active learning and reinforcement learning settings. In particular, internal adaptation appears to be very infrequent, which is most likely due to the human-centric nature of most mixed-initiative VA systems.
Adaptation in software agents could improve system usability by aligning agent responses more closely with user preferences. Internal adaptation during a session can also increase performance, while more advanced adaptive behaviors allow agents to take greater initiative in ways similar to human collaborators. Together, these mechanisms can improve both the efficiency and the quality of interaction.
\end{itemize}
\vspace{.4em}
\cbstart{\noindent\underline{\textbf{\textsc{TreeHouse} Example:}} 
However, another extension of the TreeHouse system allows the agent to apply internal adaptation and provenance tracking to assess the user’s expertise. It then adjusts its guidance style accordingly, enabling more effective navigation and more useful recommendations.}\cbend\

\vspace{.7em}
\begin{itemize}
\item[\textbf{\texttt{Opp-8}}] 
\textsc{\textbf{Human-like interaction enhances software agent comprehension, particularly in complex, unexplored system applications} | } Currently, there are only two software agents \cite{2024_TopicRefiner_HYang, 2024_Qutaber_QJiang} within the observed scope of the field that perform an interaction-type action. In contrast, most human agents perform this action type. 
Interaction capabilities in software agents could increase learnability by making agent behavior more comprehensible to humans. They could also support system testing and automation through independent human-like interaction patterns.
\end{itemize}
\vspace{.4em}
\cbstart{\noindent\underline{\textbf{\textsc{TreeHouse} Example:}} 
In an extended TreeHouse scenario, the agent first learns from user interactions, then begins mimicking the user under human guidance. Over time, this enables the agent to act as a "peer user", creating a setting much like the multi-user scenario and streamlining the exploration process.}\cbend\

\subsection{Generative Artificial Intelligence }
With the recent boom of generative AI agents within the broader computer science domain, it is also essential to address their influence within the mixed-initiative VA domain. \autoref{fig:findings_modeltypes_per_year} shows the increase per model type, where, in contrast to prior years, in 2024 and 2025, approximately 40\% of models implemented in software agents utilize generative AI. 

Generative AI has had a significant impact on the mixed-initiative VA domain. To properly explore the characteristics and benefits of generative AI, we first define what codes are most prominent in generative AI agents relative to other model types. From there, we will highlight several future development opportunities.

\subsubsection*{Characteristics of Generative AI} Currently, Generative AI is mainly utilized in the field of natural language processing and in systems without a specific domain. These agents typically generate or summarize data. Generative AI offers several advantages. First, it enhances agent awareness by relying on communication. Additionally, it improves goal awareness, as prompts can explicitly convey a user’s intent and objectives. Finally, communication chains facilitate detailed tracking of provenance through well-designed prompting strategies. This feature is particularly common in generative AI, where agents heavily rely on inputs from other agents to complete their tasks. 

However, Generative AI also has its limitations, which highlight areas for improvement and potential new implementation strategies. The generative AI agents we identified in our research generally exhibit low autonomy due to their dependence on communication-based action triggers. This reliance is evident in the observed communication patterns of incoming instructions and requests, as well as outgoing feedback and notifications. Moreover, these agents primarily function as black-box systems, which limits operational transparency. Persistence is also weak because of their reliance on single API calls and limited context windows.

\subsubsection*{Future Opportunities}
Based on current limitations, we have identified several emerging opportunities for advancing generative AI agents, including novel agent designs for mixed-initiative VA systems, new areas of exploration, and techniques to expand their capabilities.
\begin{itemize}
    \item Currently, generative AI agents are only present in a few domains -- namely, natural language processing and general data analysis -- of the VA field. Other domains will likely also benefit from the strengths of generative AI in the future. 
    \item There is a limited variety in agent roles. Currently, most generative AI agents are primarily used in Generation and Summarization, rather than in Recommendation, Labeling, and Investigation. 
\item The integration of model types capable of initiating action in mixed-initiative settings could achieve greater autonomy in generative AI. This would broaden their applications in the VA domain and reduce reliance on communication-based triggers, allowing agents to adapt through environmental cues or internal metrics.  
    \item Generative AI enhances agent and goal awareness, a capacity currently underused in mixed-initiative VA but difficult to replicate with other model types.
\end{itemize}
\section{Discussion}
In this section, we reflect on the limitations, benefits, call to actions, and future work on our design space, with the primary goal of reflecting on the created design space, and the corresponding implications. 

\subsection{Limitations}
The design space, albeit powerful, comes with certain limitations. These limitations are possible avenues to expand into future work, or in some cases are a natural consequence of design spaces.
Within the design space, we focus on intelligent agents and the infrastructure. However, this does not include a characterization of the environment, as this was not coded on a system-by-system basis due to the defined scope. 
Moreover, our design space encompasses systems that have been published in specific venues (see \autoref{sec:methodology}). While this provides some assurance regarding the quality of the papers, we also risked not capturing all existing mixed-initiative VA systems that involve at least one human and software agent when creating the design space. There may even be additional systems in industry that match our defined scope but have not been presented in any research publications.
The design space does not extend beyond the combination of VA and mixed-initiative, even though the theoretical framework that describes the intelligent agents within mixed-initiative systems could be applied to other domains outside of VA. For example, with simple adjustments to specific agent dimensions, it would be possible to describe mixed-initiative systems in different fields, such as Natural Language Processing. 
Lastly, not all papers consistently defined their system in correspondence to our design space, e.g., omitting details such as the agent model type or not exhaustively describing specific dimensions. This resulted in some viable systems being excluded or some codes being labeled through expert discussion.

\subsection{Benefits}
Through the application of the design space, a range of benefits can be conceived, as it provides a foundation of knowledge for the entire mixed-initiative VA field, encompassing existing systems, terminology, and state-of-the-art developments.
The design space allows designers and system implementers to situate their systems within the current landscape of existing works and state-of-the-art. It can also help in describing design choices during the ideation and development of novel systems. 
Moreover, the design space allows for the identification and proper understanding of existing research gaps within the field of mixed-initiative VA and the determination of future research opportunities (see \autoref{sec:opportunities}).
Our theoretical framework of the design space can be adapted to other mixed-initiative systems, such as Natural Language Processing, by making minor changes to categories and codes specific to VA.

\subsection{Call for Action to the Community}
The design space underpins the general developments and state-of-the-art for the mixed-initiative VA community. Based on our findings, we want to make several calls for action to authors within the community and to the research community as a whole.

\paragraph{Authors} We encourage authors within the community to share new and suitable systems that can be integrated into the interactive overview of the design space, ensuring that the design space remains up-to-date and provides greater benefits and contributions to the community. Furthermore, we encourage the unified description of the agent setup and its configuration as used within this paper, as mixed-initiative VA systems are becoming increasingly complex. It can also be helpful to reflect on why certain agents do or do not cover specific capabilities, such as world model sharing. Furthermore, we urge authors to be clear when defining what parts of a system are proper intelligent agents. This will consequently split intelligent agents from standard algorithms like projections that, in some systems, are used to state that mixed-initiative is prevalent. Moreover, we encourage the development of systems that have been identified as lacking within the opportunities section to broaden the domain of mixed-initiative VA. Finally, we encourage the use of a unified understanding and terminology of agent-based systems within the context of mixed-initiative VA, based on the provided design space.

\paragraph{Research Community} We call to the larger research community to create a venue for mixed-initiative systems. Moreover, it's crucial to note that new systems don't need to explore every unexplored aspect of agents within the design space at once, as more foundational work should be conducted to examine how specific agent configurations influence the mixed-initiative setting properly. 
Lastly, it is vital to value accurate reporting of system capabilities, particularly the presence and limitations of agents within mixed-initiative systems. The inherent complexity of these systems and the dynamics of human-AI collaboration are significantly amplified by their potential.  Further research is needed to systematically investigate the impact of these systems on the Human-AI interaction, ensuring that human agents can effectively leverage the benefits of mixed-initiative operation.

\section{Conclusion}
This work introduces a comprehensive design space to systematically describe the components of intelligent agents in mixed-initiative VA systems. Derived from an iterative refinement process that involves a systematic review of the literature and qualitative coding of representative systems, our design space is structured along six core dimensions: Agent Configuration + Logic, World Model, Observation, Communication, Actions, and Infrastructure. Each dimension is defined by specific categories that characterize every single agent and agent setup in the mixed-initiative system.
Our design space facilitates a structured analysis of the current research landscape on intelligent agents in mixed-initiative VA systems, enabling the identification of prevailing trends, critical challenges, and salient opportunities for future mixed-initiative VA systems. For further exploration and adoption, an interactive implementation of our design space is accessible at \href{https://dsagentmi.github.io/?p=AgentDSCHI26}{\texttt{https://dsagentmi.github.io/?p=AgentDSCHI26}}. We hope that this work will establish common ground in this burgeoning field and inspire further systems studies to evaluate the impact of the proposed design space dimensions on human-AI collaboration in mixed-initiative systems.


\bibliographystyle{ACM-Reference-Format}
\bibliography{references}

\appendix

\newpage
\section{Corpus}
In the final version of the design spaces we considered the following 90 papers: \cite{2024_CalliopeNet_QChen,2025_HEPHA_SZhou,2022_ConverseWithAnalyticChatbot_VSetlur,2022_IntentVizor_GWu,2012_InteractiveAnalysisOfBigData_JHeer,2025_InteractiveDesignOfExperiment_RSplechtna,2021_QuestionComb_RSevastjanova,2025_RSVPForVPSA_MKlaffenboeck,2025_Sprout_YLiu,2022_SupportingSerendipitousDiscovery_MJasim,2022_VideoModerator_TTang,2025_VisCars_PMoens,2012_VAAgentBasedFrameworkForIntrutionAlertAnalysis_RShittu,2025_WhatIF_AMishra,2021_QualitativeCausalModeling_FHusain,2019_UserbasedVAWorkflowForEMA_DCashman,2025_WizzardGuidanceStrategiesAndDynamics_FSperrle,2021_ArchiText_HKim,2010_Articulate_YSun,2019_AugmentingVisualizationsWithInteractiveDataFacts_ASrinivasan,2007_ContextAwareAdaptiveInfromationRetrieval_ZWen,2024_Dash_DBromley,2023_Datatales_NSultanum,2022_DeconstructingCategorizationInVisualizationRecommendation_DJung,2024_DiscipLink_CZheng,2023_DiverseInterationRecommendationForPublicUsers_YLi,2020_DivingInsights_AMcNutt,2020_explAIner_TSpinner,2025_GuidanceSourceMatters_ANarechania,2024_GuidedByAI_SHa,2024_GuidedVAForImageSelectionInTimeAndSpace_IPerezMessina,2024_InkSight_YLin,2025_InsigHTable_GLi,2006_InteractiveVisualSynthesisOfAnalyticalKnowledge_DGotz,2024_InterpretingHighDimensionalProjectionsWithCapacity_YZhang,2021_InterwarvingMultimodalInteraction_ASrinivasan,2024_JailbreakLens_YFeng,2025_KMTLabeler_HWang,2021_LearningContextualizedUserPreferences_FSperrle,2025_Leva_YZhao,2025_LightVA_YZhao,2021_MI3_YZhang,2015_MixedInitiativeVAUsingTaskDrivenRecommendations_KCook,2025_MixingLinters_AMcNutt,2022_MutliVision_AWu,2010_PredictiveAnalyticsUsingBlackboard_JYu,2017_PrivacyPreserving_JKChou,2021_Snowy_ASrinivasan,2024_Socrates_GWu,2024_SupportingGuidedEVAOnTimeSeriesData_YShi,2024_TheDataSaysOtherwise_YFu,2007_Tibor_DLiu,2024_TopicRefiner_HYang,2022_TowardsVisualExplainableActiveLerning_SJia,2019_Viana_FSperrle,VideoPro_He_2024,2024_VisStoryMaker_DJackFreireBraga,2019_VAForTopicModeling_MElAssady,2020_VisualInteractionWithDLModelsThroughCollaborative,2016_Voyager_KWongsuphasawat,2023_Causeworks_TKapler,2024_Qutaber_QJiang,2025_Dango_WHChen,2024_iToT_ABoyle,2020_AnchorViz_JSuh,2022_ChartSeer_JZhao,2024_ConceptInduction_MSLam,2023_ConceptEVA_XZhang,2023_DASSGood_AWentzel,2025_Divisi_VSivaraman,2019_DoWhatIMean_MTroy,2019_EvaluatingPreferenceCollection_CKuhlman,2016_Eviza_VSetlur,2020_HumanSteerableAI_AGonzalezMartinez,2019_Fairvis_AACabrera,2021_GeonoCluster_SDas,2011_GuidingFeatureSubsetSelection,2022_HowDoAlgorithmicFairness_CRares,2021_IntegratingPriorKnowledge_APister,2021_LayoutExOmize_PSchader,2018_Podium_EWall,2025_ScholarMate_RYe,2015_SeeDB_MVartak,2020_SemanticConceptSpaces_MElAssady,2019_Sherpa_ZCui,2024_Stile_SKabir,2024_Talk2Data_YGuo,2016_TimeFork_SKarthik,2018_VisIRR_JChoo,2019_VASystemForSubjectMatter_CHagerman} 

\section{Scope of considered Work}
\label{appendix:exampleExclusion}
Within this section, multiple examples of works that were excluded during the \textit{Filtering} phase (\autoref{sec:methodology}) will be provided.

\subsection{Not Visual Analytics} \citet{Ying2023MetaGlyph} describe a mixed-initiative system that generates metaphoric glyph-based visualizations from a spreadsheet in order to provide visual metaphors for users. The authors implement a mixed-initiative system, and at least one human and one software agent can be identified. However despite fitting the mixed-initiative criteria, the domain of the paper is information visualization rather than visual analytics. As the system does not fall under visual analytics (\autoref{sec:ScopeAndDefinitions}), it has been excluded from the literature corpus.

\subsection{Not Mixed-Initiative} Throughout "Data Brushes: Interactive Style Transfer for Data Art" \cite{Dubey2019DataBrushes}, the concept of data brushes are described. These use neural style transfer models to alter images according to specific styles. A certain content image is transformed through a specific brush, resulting in a specific composition. The system involves a human agent that applies specific brushes. Although there is no software component that can be identified as an intelligent agent (Definition in \autoref{sec:ScopeAndDefinitions}) within the system, as the human agent is simply interacting with the environment and changing the composition using different brushes. Thus, based on our definition of mixed-initiative (\autoref{sec:ScopeAndDefinitions}) and the requirement of including at least one human agent and one software agent, the paper was excluded.

\subsection{A Robotic Agent} The work by \citet{Bohus2009DialogOpenWorld}, details a mixed-initiative (non-visual analytics) dialog system involving a human actor and a robotic agent. The robotic agent is able to observe the environment and recognize other (human) agents within said environment. Human agents can then begin a trivia game with the robotic agent. However, in addition to the system not being within the visual analytics domain, the described robotic agent is physical in nature, and our scope focuses only on human, wizard, and software agents. Thus, this paper was not considered within the scope of this work. 

\subsection{No System Described} \citet{Ceneda2017CharacterizingGuidance} describe a framework for guidance inside the visual analytics domain, where guidance can be a form of mixed-initiative. The paper details multiple components and even provides examples of guidance-based systems and how the framework would apply in these cases. However, no single system is described inside the paper with sufficient detail for it to be coded within the design space. Thus, this paper was excluded from the literature corpus.

\section{Use Case Scenarios}
\label{appendix:usecases}
This section presents two compelling use cases that demonstrate the practical application of our design space. These scenarios illustrate the design space's adaptability and value for diverse stakeholders -- including researchers and system designers -- while highlighting the complex relationships and trade-offs among different dimensions and codes.

\paragraph{Active Learning Scenario} A research team from an intelligence company wants to create a system to classify whether a specific image contains a threat, such as a weapon. However, the company lacks the necessary training data to develop such a classification model. Based on the design space, they identify existing systems that use active learning methods, which would solve their issue. Using the design space, they choose to give their \underlineColor{colorConfigurationLogic}{Semi-Supervised} model a sense of agency. Allowing it to rank data points as a (\underlineColor{colorConfigurationLogic}{Ranker}) and \underlineColor{colorCommunication}{Request} for labels from the human agent.
Although active learning approaches achieve high gains with minimal user feedback, the company aims to integrate additional methods to improve the quality of the labeling process further. Evidence from outside the VA domain indicates that the infrastructure can be adapted to support multiple software agents working in a \underlineColor{colorInfrastructure}{Competitive} setup with one another.
Therefore, they extend their system to include multiple \underlineColor{colorConfigurationLogic}{Semi-Supervised} classification agents that compete for the best performance. Using the proposed design space, the intelligence company was able to examine existing systems within the field of mixed-initiative VA to identify a suitable system. Moreover, by critically examining the different dimensions, they were able to identify a novel mode of agent interplay that facilitates better performance for their use case.

\paragraph{Exploration Scenario} A team of digital designers aims to develop an in-house tool for new, up-and-coming generative AI tools that facilitate the creation of advertising artworks, such as logos or banners. The team wants to ensure that they have complete control of the creative decision-making process, such that employees can roll back decisions and see alternative options by creating a tree of generated responses, splitting over alternative prompts, and growing deeper over follow-up responses. The system includes one human agent and one software agent, which uses the \underlineColor{colorConfigurationLogic}{Generative AI} model type through implementing modern language models. To properly implement the system, the digital designers consider the world model dimension from the design space. They conclude that the software agent should have proper \underlineColor{colorWorldModel}{Provenance Tracking}, allowing the agent to keep track of the \underlineColor{colorWorldModel}{Observations} it has made of the artworks and prompts.
Additionally, the team decides that designers should be able to re-open existing sessions, and the agent thus requires \underlineColor{colorWorldModel}{Continuous Persistence}. Within the design space, the designers notice that their current approach is lacking \underlineColor{colorCommunication}{World Model Sharing} by the \underlineColor{colorConfigurationLogic}{Generative AI} model type, which makes it hard for the human agent to understand the software agent's decisions. They configure the responses such that the software agent provides context to the observations of the prompt and the generated response. By utilizing the framework provided by the design space, the team can effectively define the scope of their system and explore innovative ideas. 

\newpage
\section{Definitions of Categories and Codes}
\label{appendix:definitions}

\subsection{Configuration and Logic}

    \begin{xltabular}{\textwidth}{lX}
    \caption{Definitions of the Categories and Codes in the Configuration and Logic dimension}
    \\\toprule
 \multicolumn{2}{X}{\underlineColor{colorConfigurationLogic}{Configuration and Logic}}\\\midrule
 \multicolumn{2}{X}{\textbf{Agent Type}}\\
 \multicolumn{2}{X}{\textit{What is the type of agent, are they a human, a wizard or software?}}\\
          Human& A form of agent that manifests itself as a physical human.\\
          Wizard& A form of agent often found in 'Wizard of Oz' studies. The agent manifests itself as a physical human but takes the responsibilities of a software agent. \\
 Software&A form of agent that describes an encapsulated part of a system that fits the criteria of an agent. It manifests itself as a digital entitiy.\\ \bottomrule

 \multicolumn{2}{X}{\textbf{Agent Role}}\\
 \multicolumn{2}{X}{\textit{What are the roles that an agent takes within a given system?}}\\
 Analyzer& An agent takes the role of observing data present within the environment to draw internal conclusions.\\
 Instructor& An agent takes the role of providing instructions to other agents in order to drive a shared workflow forward.\\
 Interactor&An agent interacts with the UI and infrastructure layer to bring about changes.\\
 Recommender& An agent provides recommendations, suggestions or feedback to other agents within the system.\\
 Explorer& An agent takes the role of skimming through data present within the environment, to create an understanding of the data present.\\
 Visualizer& An agent creates visualizations based on the data presented in the environment.\\
 Generator& An agent generates new data and puts this into the environment. This excludes visualizations, recommendations, rankings and summarizations.\\
 Ranker& An agent ranks data points within the environment according to internal criteria, and creates a ranking based on this.\\
 Judge& An agent takes the role of judging data present within the environment for validity, precision or accuracy.\\
 Summarizer& An agent condenses multiple data points from the environment into a smaller format, to describe all data points with fewer information (i.e., a Summary).\\
 Labeler& An agent takes on the role of labeling data, to provide, based on features, a suitable tag or label for a specific data point from the environment.\\\bottomrule
 \multicolumn{2}{X}{\textbf{Model Type}}\\
 \multicolumn{2}{X}{\textit{What learning techniques are used by an agent for thinking and decision-making?}}\\
 Rule-Based& 
Rule-Based models utilize basic mathematical operations and algorithms in their decision-making process.\\
 Supervised& Supervised models utilize labels in their training and decision-making process.\\
 Unsupervised& Unsupervised models utilize un-labeled data in their training and decision-making process.\\
 Semi-Supervised& Semi-supervised models combine un-labeled data and labeled data in their training and decision-making process.\\
 Reinforcement Learning& Reinforcement learning models utilize adaption to train and make decisions based on interactions (actions or observations) within the environment.\\
 Generative AI& Generative AI extends to any sort of complex models (e.g., LLMs) that use previously trained embeddings in order to generate new responses.\\
 Human Brain& The human brain represents a complex decision-making organ, pertaining to users, wizards and more generally human agents.\\\bottomrule
 \multicolumn{2}{X}{\textbf{Autonomy}}\\
 \multicolumn{2}{X}{\textit{Is the agent able to make decisions and act by itself or does it require external stimuli?}}\\
 Internal Deliberate& Actions are decided based on internal deterministic logic or interactions with the environment such as observations or actions.\\
 Internal Spontaneous& Actions are decided based on an internal random seed encapsulated within the agent itself. \\
 External& Actions are induced by communication from other agents or through the configuration layer (i.e., the agent reacts to other agents and is not autonomous).\\\bottomrule
 \multicolumn{2}{X}{\textbf{Adaptation}}\\
 \multicolumn{2}{X}{\textit{In what way is the agent adapting within the system?}}\\
 Internal Adaptation& The agent adapts based on internal logic and their world model.\\
 External Observations& The agent adapts based on external observations of the environment or the infrastructure.\\
External Communicaiton& The agent adapts based on signals from other agents or the configuration layer.\\
 None& The agent employs no adaptation.\\
 \bottomrule
    \label{tab:DefinitionsConfigurationLogic}
 \end{xltabular}

\subsection{World Model} 
\begin{xltabular}{\textwidth}{lX}
 \caption{Definitions of the Categories and Codes in the World Model dimension}
\\\toprule
 \multicolumn{2}{X}{\underlineColor{colorWorldModel}{World Model}}\\\midrule
 \multicolumn{2}{X}{\textbf{Provenance Tracking}}\\
 \multicolumn{2}{X}{\textit{What types of historical records are part of the agent’s world model?}}\\
          Observation History& The world model includes information on the observations done by the agent on the environment.\\
          Action History& The world model includes information on the actions taken by the agent on the environment.\\
 Interaction History& The world model includes information on series of actions taken by other agents within the environment.\\
 Communication History& The world model includes information on communication between the agent and other agents.\\
 None&The world model does not capture any history.\\ \bottomrule
 \multicolumn{2}{X}{\textbf{Persistence}}\\
 \multicolumn{2}{X}{\textit{To what degree is the World Model persistent?}}\\
 Continous& The world model is persistent within the agent and is retained even if the current session is ended (i.e., The world model is stored in long-term memory).\\
 Session-Based& The world model is persistent within the agent during the duration of a session (i.e., the world model is stored in short-term memory).\\
 Not Persistent& The world model is not persistent (i.e., the world model is a set or subset of the most recent observations).\\\bottomrule
 \multicolumn{2}{X}{\textbf{Data Awareness}}\\
 \multicolumn{2}{X}{\textit{What is the degree of the agent’s awareness of data present within the environment?}}\\
 Full Data Awareness& 
The agent is aware of the full environment, this excludes the infrastructure layer.\\
 Aligns with Observations& The agent is aware of a subset of the environment, in-line with all their observations.\\
 Subpart of Observation& The agent is aware of a subset of the environment, and their awareness extends to a subset of their observations.\\\bottomrule
 \multicolumn{2}{X}{\textbf{Agent Awareness}}\\
 \multicolumn{2}{X}{\textit{Is the agent aware of other agents within the system?}}\\
 Aware of all Agents& The agent is aware of all the other agents within the system.\\
 Aware of a Subset of Agents& The agent is only aware of a subset of the other agents within the system.\\
 Not Aware of others& The agent is not aware of other agents within the system.\\\bottomrule
 \multicolumn{2}{X}{\textbf{Task Awareness}}\\
 \multicolumn{2}{X}{\textit{Is an agent aware of its own tasks, and is the agent aware of the tasks of other agents?}}\\
 Self Aware& The agent is aware of its task, and is able to reason about it and make decisions based on said task.\\
 Aware of other Agents Tasks& The agent is aware of the task of at least one other agent within the system domain.\\
No Awareness& The agent is not aware of its own task and that of other agents.\\
\bottomrule
    \label{tab:DefinitionsWorldModel}
\end{xltabular}

\subsection{Observations}

    \begin{xltabular}{\textwidth}{lX}
     \caption{Definitions of the Categories and Codes in the Observations dimension}
    \\\toprule
 \multicolumn{2}{X}{\underlineColor{colorObservation}{Observations}}\\\midrule
 \multicolumn{2}{X}{\textbf{Data Structure}}\\
 \multicolumn{2}{X}{\textit{When observing the environment, how is the observed data structured?}}\\
          Structured& The observed sensory data is structured, e.g., a table.\\
          Semi-Structured& The observed sensory data is semi-structured, e.g., JSON-format.\\
 Unstructured& The observed sensory data is unstructured, e.g., text.\\ \bottomrule
 \multicolumn{2}{X}{\textbf{Sensor Modality}}\\
 \multicolumn{2}{X}{\textit{What sensory modalities are part of the agent’s observation?}}\\
 Text& The agent observes textual data through their observations, textual data is represented through Strings and Character values.\\
 Numeric& The agent observes numerical values through their observations, numerical value can be represented as Integers, Floats, but also as Vectors or Matrices.\\
 Visual& The agent observes visual data through their observations, which is numeric but represents an image and/or visualization.\\
 Code& The agent observes code-based data through their observations, this data is textual but represents a set of instructions that can be executed by an agent.\\\bottomrule
 \multicolumn{2}{X}{\textbf{Context Level}}\\
 \multicolumn{2}{X}{\textit{What part of the environment is observed by the agent?}}\\
 One Input& 
The agent observes a single unique input from the environment.\\
 Multiple Inputs& The agent observes at least two unique inputs from the environment.\\
 Whole Environment& The agent observes the full environment.\\\bottomrule
 \multicolumn{2}{X}{\textbf{Synchronism}}\\
 \multicolumn{2}{X}{\textit{Is the environment able to change between an agent observing and consequently acting based on said observation?}}\\
 Synchronized& The agent observes the environment with certainty, the state of the environment that is being observed is synchronized. Actions and communication based on an observation take place before the environment updates.\\
 Asynchronized& The agent observes the environment with a degree of uncertainty, the state of the environment that is being observed is asynchronized. Actions and communication based on an observation happen with no guarantee that the environment has updated.\\\bottomrule
 \multicolumn{2}{X}{\textbf{Trigger}}\\
 \multicolumn{2}{X}{\textit{In what way is the agent adapting within the system?}}\\
 Internal& The observation starts due to internal logic and metrics.\\
 Communication Based& The observation starts due to external signals from other agents.\\
Infrastructure Based& The observation starts due to external signals from the infrastructure layer.\\
\bottomrule
    \label{tab:DefinitionsObservations}
\end{xltabular}

\subsection{Communication}
\begin{xltabular}{\textwidth}{lX}
 \caption{Definitions of the Categories and Codes in the Communication dimension}
\\\toprule
 \multicolumn{2}{X}{\underlineColor{colorCommunication}{Communication}}\\\midrule
 \multicolumn{2}{X}{\textbf{Payload}}\\
 \multicolumn{2}{X}{\textit{Is there any form of communication from an agent, and does the signal include a payload?}}\\
          Signal& The agent communicates directly to another agent and exclusively provides instructions without sending data (outside of the signal) into the environment.\\
          Signal and Data& The agent communicates directly to another agent and provides both instructions and data.\\
 None& The agent does not communicate with other agents.\\ \bottomrule
 \multicolumn{2}{X}{\textbf{Type Outgoing}}\\
 \multicolumn{2}{X}{\textit{What are the signal types of the outgoing communication from an agent?}}\\
 Notification& The agent informs another agent.\\
 Instruction& The agent gives instructions to another agent.\\
 Request& The agent requests information from another agent.\\
 Feedback& The agent provides feedback to another agent.\\
 Availability& The agent provides its availability (e.g., Busy or Available) to another agent. \\
 None& The agent does not communicate with other agents.\\\bottomrule
 \multicolumn{2}{X}{\textbf{Type Incoming}}\\
 \multicolumn{2}{X}{\textit{What are the signal types of the incoming communication to an agent?}}\\
 Notification& 
The agent gets informed by another agent.\\
 Instruction& The agent gets instructions from another agent.\\
 Request& The agent gets requested information by another agent.\\
 Feedback& The agent is provided feedback by another agent.\\
 Availability& The agent receives availability (e.g., Busy or Available) updates from another agent.\\
 None& The agent does not receive communication from other agents.\\\bottomrule
 \multicolumn{2}{X}{\textbf{Trigger}}\\
 \multicolumn{2}{X}{\textit{What trigger type is initiating the agent-to-agent communication?}}\\
 Internal& The communication starts due to the agent's internal logic and metrics.\\
 Context Based& The communication starts based on the agent's observations.\\
 Communication Based& The communication starts due to external signals from other agents.\\
 Infrastructure-based, & The communication starts due to external signals from the infrastructure layer.\\
 None& There is no explicit timing to communication.\\\bottomrule
 \multicolumn{2}{X}{\textbf{Internal Sharing}}\\
 \multicolumn{2}{X}{\textit{Is an agent sharing its world model and logic with another agent?}}\\
 Operational Sharing& The agent communicates their operational representation (e.g., internal metrics or methods) with at least one other agent.\\
 World Model Sharing& The agent communicates their world model with at least one other agent.\\
 None& The agent does not provide a status representation to other agents.\\
 \bottomrule
    \label{tab:DefinitionsCommunication}
 \end{xltabular}

\subsection{Actions}
\begin{xltabular}{\textwidth}{lX}
\caption{Definitions of the Categories and Codes in the Actions dimension}
\\\toprule
 \multicolumn{2}{X}{\underlineColor{colorActions}{Actions}}\\\midrule
 \multicolumn{2}{X}{\textbf{CRUD Type}}\\
 \multicolumn{2}{X}{\textit{What is the type of action taken by an agent?}}\\
          Create& Create relates to any new information (or data) that is being added to the environment. \\
          Update& Update relates to any information that is adjusted/manipulated within the environment. \\
 Delete& Delete relates to any information that is removed or obscured from the environment.\\ \bottomrule
 \multicolumn{2}{X}{\textbf{Type/Modality}}\\
 \multicolumn{2}{X}{\textit{What is the modality of the action taken by an agent within the environment?}}\\
 Writing& The agent creates, updates or deletes textual information within the environment.\\
 Counting& The agent creates, updates or deletes numerical information within the environment.\\
 Coding& The agent creates, updates or deletes some information that represents instructions to other agents or instructions to itself.\\
 Visualizing& The agent creates, updates or deletes visual information within the environment that can be observed within the UI.\\
 Interacting& The agent is manipulating and utilizing tools within the UI, including API calls.\\\bottomrule
 \multicolumn{2}{X}{\textbf{Affected Scope}}\\
 \multicolumn{2}{X}{\textit{What part of the environment and world model is effected by an action?}}\\
 Entire Environment& 
The action impacts the whole environment. \\
 Part of World Model& The action impacts only a subset of the environment and the action is inside of the agent's world model.\\
 Outside of World Model& The action impacts only a subset of the environment and the action is outside of the agent's world model.\\\bottomrule
 \multicolumn{2}{X}{\textbf{Trigger}}\\
 \multicolumn{2}{X}{\textit{What triggers an agent’s action to happen?}}\\
 Internal& The action starts due to internal logic and metrics.\\
 Context Based& The action starts based on the agent's observations.\\
 Communication Based& The action starts due to external signals from other agents.\\\bottomrule
 \multicolumn{2}{X}{\textbf{Delay/Timing}}\\
 \multicolumn{2}{X}{\textit{When an agent takes an action, is the action delayed or executed immediately?}}\\
 No Delay& The action is immediately executed upon the environment after it has been triggered.\\
 Delay with Static Duration& The action is excuted with a pre-determined delay after it has been triggered.\\
Delayed Based on a Metric& The action is executed once specific metrics are met after it has been triggered.\\
 Delayed Until Action Acknowledged by Other Agent& The action is executed upon acknowledgement from another agent after it has been triggered.\\
 \bottomrule
    \label{tab:DefinitionsActions}
 \end{xltabular}

\subsection{Infrastructure}
    \begin{xltabular}{\textwidth}{lX}
    \caption{Definitions of the Categories and Codes in the Infrastructure dimension}
    \\\toprule
 \multicolumn{2}{X}{\underlineColor{colorInfrastructure}{Infrastructure}}\\\midrule
 \multicolumn{2}{X}{\textbf{Initialized Modules}}\\
 \multicolumn{2}{X}{\textit{What modules are initialized within the infrastructure?}}\\
 Configuration Module& The infrastructure layer and all agents are initialized.\\
          Observation Module& The infrastructure layer initializes an observation module within all agents, which defines what observations they are able to make.\\
          Action Module& The infrastructure layer initializes an action module within all agents, which defines what actions they are able to take.\\
 Communication Module& The infrastructure layer initializes a communication module within the system, which defines what agent-to-agent communication is facilitated.\\ \bottomrule
 \multicolumn{2}{X}{\textbf{Dynamic Modules}}\\
 \multicolumn{2}{X}{\textit{What models are dynamically configured by the infrastructure layer during run-time?}}\\
 Configuration Module& The infrastructure layer dynamically changes the initialization of agents.\\
 Observation Module& The infrastructure layer dynamically changes the observation module within at least one agent, changing what observations they are able to make.\\
 Action Module& The infrastructure layer dynamically changes the action module within at least one agent, changing what actions they are able to take.\\
 Communication Module& The infrastructure layer dynamically changes the communication module of the system, changing the way agent-to-agent communication is facilitated, or changing which agents are communicating with one another.\\
 None& The infrastructure layer does not dynamically change any models.\\\bottomrule
 \multicolumn{2}{X}{\textbf{Trigger of Dynamic Modules}}\\
 \multicolumn{2}{X}{\textit{What triggers the dynamic adjustment of the modules?}}\\
 Planned Time Steps& 
Dynamic module updates are executed after pre-defined and scheduled time steps. \\
 Based on Agents& Dynamic module updates are executed based on agent actions, and world model.\\
 Based on Environment& Dynamic module updates are executed based on the state of the environment.\\
 None& Dynamic module updates do not happen.\\\bottomrule
 \multicolumn{2}{X}{\textbf{Agent Interplay}}\\
 \multicolumn{2}{X}{\textit{What is the nature of the relationship of the interplay between agents within the mixed-initiative VA
system?}}\\
 Independent& The agents take part in individual tasks, based on their own merit, and do so independently.\\
 Cooperative& The agents collaborate in a shared task, based on shared effort, they achieve a specific result, directly contributing to the same task.\\
 Competitive& The agents compete towards a shared goal or task, based on competition, they can achieve a specific result.\\
 \bottomrule
    \label{tab:DefinitionsInfrastructure}
 \end{xltabular}

\end{document}